\documentclass{article}
\usepackage[utf8]{inputenc}
\usepackage{float}
\usepackage{algcompatible}
\usepackage{authblk}
\usepackage{xcolor}
\usepackage[T1]{fontenc}
\usepackage{amsmath}
\usepackage{algpseudocode,algorithm}
\usepackage{algorithmicx}
\usepackage{indentfirst}
\usepackage{enumerate}
\usepackage{booktabs}
\usepackage{graphicx}
\usepackage{placeins}
\usepackage[font=small,bf,justification=centering]{caption}
\usepackage{subcaption}
\usepackage[export]{adjustbox}
\usepackage{multirow}
\usepackage{array}
\usepackage[symbol]{footmisc}
\usepackage{amssymb}
\usepackage{txfonts}
\usepackage{pxfonts}
\usepackage{bbm}
\usepackage[makeroom]{cancel}
\usepackage{relsize}
\usepackage{enumitem}
\usepackage{verbatim} 
\usepackage{titlesec}
\usepackage{fancyhdr}
\DeclareMathOperator*{\argmax}{arg\,max}
\DeclareMathOperator*{\argmin}{arg\,min}
\renewcommand{\binom}[2]{\left(\genfrac{}{}{0pt}{}{#1}{#2}\right)}

\newtheorem{thm}{Theorem}
 
 \newtheorem{lem}[thm]{Lemma}

\newtheorem{definition}{Definition}[section]

\title{Probabilistic Context Neighborhood Model for Lattices}

\author{ Débora  F. Magalhães \\ 
		\small{Department of Statistics, UFMG, Brazil, { \it e-mail} $debora_fmag@hotmail.com$ }\\ 

\and Aline M. Piroutek \\
		\small{Department of Statistics, UFMG, Brazil, { \it e-mail}: lyne.piroutek@yahoo.com.br  }\\
  
  \and Denise Duarte \\ 
	\small{Department of Statistics, UFMG, Brazil, \it e-mail: denised@ufmg.br}
\and Caio  Alves\\
\small{Alfred Rényi Institut of Mathematics, Hungary,  { \it e-mail} caiotmalves@gmail.com}
}

\date{ }

\begin{document}

\maketitle

\begin{abstract}

    We present the Probabilistic Context Neighborhood model designed for two-dimensional lattices as a variation of a Markov Random Field assuming discrete values. In this model, the neighborhood structure has a fixed geometry but a variable order, depending on the values of the neighbors. Our model extends the Probabilistic Context Tree model, originally applicable to one-dimensional space, and retains its advantageous properties, such as representing the dependence neighborhood structure as a graph in a tree format, facilitating the understanding of model complexity. Furthermore, we adapt the algorithm used to estimate the Probabilistic Context Tree to estimate the parameters of the proposed model. We illustrate the accuracy of our estimation methodology through simulation studies. Additionally, we apply the Probabilistic Context Neighborhood model to spatial real-world data, showcasing its practical utility. 
    

\end{abstract}

\section{Introduction}
\label{sec:intro}

A Markov random field (MRF) is a type of model used to explain how data interacts with one another \cite{KINDERMANN1980}, \cite{BESAG1975}. The MRF framework conditions the probability of a random variable on its neighbors, which is based on the well-known Markovian property. This type of model is very versatile and can be used to model time dependence, spatial dependence, and even space-time dependence in various applications.   
    
    One of the main applications of the MRF methodology is image analysis and remote sensing \cite{GEMAN1984}. Understanding the interactions between pixels can help recover \cite{KIM1995}, segment \cite{WU2016}, synthesize and correctly classify images  \cite{SUBUDHI2014}, \cite{ZHANG2017}. MRF models are not limited to computer vision and geostatistics applications; they can also be used in biology to model gene interactions.  For instance, MRF-based procedures have been used to identify subnetworks related to breast metastasis or death from breast cancer \cite{WEI2007}. In the field of neuroscience, MRF has been used to study brain development and how different regions of the brain are impacted by neighboring regions and time \cite{LIN2015}. MRF models can also be applied to economics to study the interactions between individuals, households, and financial institutions, as demonstrated in \cite{ONURAL2021}. In a study by Fahrmeir (2001), a different approach was taken to examine the impact of districts in Germany on their unemployment rates. The study made use of Markov Random Fields (MRF) to investigate the spatial effect.  Social networks are another area where MRF has gained popularity. In  \cite{WEST2018}, MRF was used to model person-to-person interactions, taking into account the overall social network structure and sentiment analysis. A neighboring profiles-based MRF method is used for recommending new users or items in commercial applications, as presented in \cite{PENG2016}.  The MRF model has many applications, as detailed in \cite{KINDERMANN1980} and \cite{LEMUS2021}. For the specific case of Gaussian Markov Random Fields, please refer to \cite{RUE2005}.
    
    Our research focuses on examining the spatial dependence of MRF processes in two-dimensional lattices with discrete values, particularly in $\mathbb{Z}^2$. We propose the Probabilistic Context Neighborhood (PCN) model, which utilizes a tree representation to depict the MRF's spatial dependence on lattices in $Z^2$, similar to the Probabilistic Context Tree (PCT) model for one-dimensional discrete framework proposed by Rissanen in 1983. The PCN model allows the neighborhood's order to vary from one site in the lattices to another. The purpose of this model
is to provide insight into the dependency of sites on their neighbors through learning the dependency structure and estimating the conditional probabilities that determine the value of a site. As shown in \cite{FRANK1986}, assumptions about the dependency structure of a graph can lead to various modeling strategies. However, unlike the graphs considered in their work, the graphs we consider are not random. The PCN model evaluates the interaction of lattices with a fixed structure of nodes and edges. The randomness lies in the tree dependency structure and its conditional probabilities. It is important to note that the key difference between a discrete  MRF with a fixed order neighborhood and a PCN lies in the fact that for the former, the order of dependence between each site's neighboring elements is predetermined and known, whereas, in the latter, it is not. Then, the estimation procedure must also learn the PCN's dependence order and the contexts. This fact makes the estimation procedure much more complicated.

   Estimating the parameters of an MRF is typically done using potentials (as stated in \cite{LI2001}), but this approach does not apply to the PCN model. The PCN model has a specification that provides the probability of a site based on its neighborhood configuration. The neighborhood configuration that determines a site's conditional probability is called its "context," and the size of this context can vary from site to site. We will explain this definition in more detail later on.  In \cite{LOCHERBACH2011}, the authors present a consistent estimator for the radius of the smallest ball containing the context. They also provide an algorithm to calculate this estimator and yield an explicit upper bound for the probability of wrong estimation.
   
 The article \cite{CSISZAR2006b} introduces the pseudo-Bayesian information criterion ({PIC}) as a tool for choosing models. PIC can help identify a reliable estimator for the smallest region that defines the conditional probability of an MRF. However, the authors did not give practical instructions on calculating this estimator, which remains an open question. The PCN model provides a solution for lattices in $\mathbb{Z}^2$ by consistently estimating the source's dependency structure, given a sample. Unlike previous models that directly estimate the minimal neighborhood, the PCN model assumes a specific context geometry, allowing for a direct analogy with the Probabilistic Context Tree (PCT) presented in \cite{RISSANEN1983}. This analogy enables us to represent dependency structures using a graph in a tree format, similar to a PCT, making the estimation process more manageable. We have developed an algorithm to estimate "context neighborhoods" from a given sample. This algorithm is combined with a modified pruning procedure for the one-dimensional case, as proposed in \cite{CSISZAR2006a}. The resulting algorithm enables a relatively fast and straightforward implementation of the PCN model.
       
    We present a simulation study of the dependency structure of a process, assuming the existence of an underlying MRF. Our simulation results for black and white images demonstrate the effectiveness of our proposed algorithm in accurately recovering the dependence structure that generates the process.

    We demonstrate a practical application of this methodology to a real-world dataset. Considering the alarming number of fire outbreaks in the Pantanal Biome of the Center-West Region of Brazil, particularly in September 2020 \cite{INPEfocos}, we have conducted a study to determine the spatial correlation of fires in that region.
    
    Our work is structured in the following way. Section 2 briefly introduces important concepts and results that form the basis of the PCN model presented in Section 3. We present a simulation study and its results in Section 4. In Section 5, we examine the spatial dependence of fires in the Pantanal biome in Brazil that occurred in September 2020. Finally, we conclude with our final thoughts in Section 6.

    \section{Background and Motivation} 
\label{sec:2}

    We present in this section a few methodologies that address (at some capacity) the problem of parameter estimation and model selection in the Markov framework. First,  we introduce the concept of a Markov random field and a few existing results related to it. Our aim is to show what has been proposed, but also the gaps left unresolved which the PCN model seeks to fill.

\subsection{Markov Random Fields (MRFs)}
\label{sec:mrf}

    Let us now consider the general case of a $d$-dimensional lattice $\mathbb {Z}^d$. The points $ i \in \mathbb {Z}^d$ are called sites. The cardinality of a set $\Delta \subset \mathbb{Z}^d $ is denoted as $|\Delta|$. We denote by $\Subset$ and $\subset$ the inclusion and strict inclusion, respectively. Subsets of $\mathbb Z^d$  will be denoted by uppercase Greek letters. Thus, if $\Lambda$ is a finite set of sites, then $\Lambda \Subset \mathbb Z^d$.

    A random field is a family of random variables indexed by the site $ i $ of a lattice, $\{X(i): i \in \mathbb Z^d \} $, where each $X(i)$ is a random variable that takes values in a finite alphabet $A$. We denote the set of all configurations of the random field as $\Omega=A^{\mathbb Z^d}$. For realizations of $X(\Delta)$, we use the notation $a(\Delta)=\{a(i) \in A:i\in \Delta \}$.

    The joint distribution of $X(i)$ is given by:
    \[Q \big( a(\Delta) \big) = P\big( X(\Delta) = a(\Delta) \big),\] for $\Delta \subset \mathbb Z^d $ and $a(\Delta) \in A^{\Delta}$.

    And the  conditional probability is defined by:
    \[ Q \big(a(\Delta) \,\big| \, a(\Phi) \big) = P\big (X (\Delta) = a (\Delta) \, \big| \, X (\Phi) = a(\Phi)\big)\]
    \noindent for all disjoint regions $\Delta$ and $\Phi$ where $Q(a(\Phi))>0$ .

    We say that the process is a Markov random field (MRF) if there exists a neighborhood $\Gamma_i$, satisfying for every $i \in \mathbb Z^d$
    \begin{align}
    \label{eq:mrf}
    P \big(X (i) = a (i) \, \big| \, X (\mathbb Z^d \backslash i) = a (\mathbb Z^d \backslash i) \big) = P \big( X(i) = a (i) \, \big| \, X (\Gamma_i) = a (\Gamma_i) \big),
    \end{align}
    \noindent where a neighborhood $\Gamma_i$ (of the site $i$) means a finite, central-symmetric set of sites with $i \notin \Gamma_i$.

\subsubsection{Variable-neighborhood Random Field (VNRF)}
\label{subsec:vnrf}

    If estimating a Markov chain can be challenging as the order dependency grows, the problem of estimating the parameters of an MRF is much more complicated   even when the size of the neighborhood is known. In an attempt to minimize this issue, the variable-neighborhood random field ({VNRF}) model was created in \cite{LOCHERBACH2011}, generalizing to random fields in $\mathbb{Z}^d$ the concept of a Probabilistic Context Tree (PCT) introduced by \cite{RISSANEN1983} for one-dimensional case. The PCT model is a Markov chain depending on a variable length of lagged values. The relevant past that influences the next outcome is called \textit{context}. A context may be short or long depending on the length of the string needed to determine the conditional probability of the next symbol. By only storing the minimal states, there is a reduction in the number of parameters in a PCT model compared to a full-order Markov chain. The set of all contexts (allowed to be of variable length) was represented as the set of leaves of a rooted tree. The root of the tree represents the present state of the chain, while the leaves represent past states as we move down the tree.
    
      \begin{figure}[H]
    \centering
    \includegraphics[height=21em, width=42em]{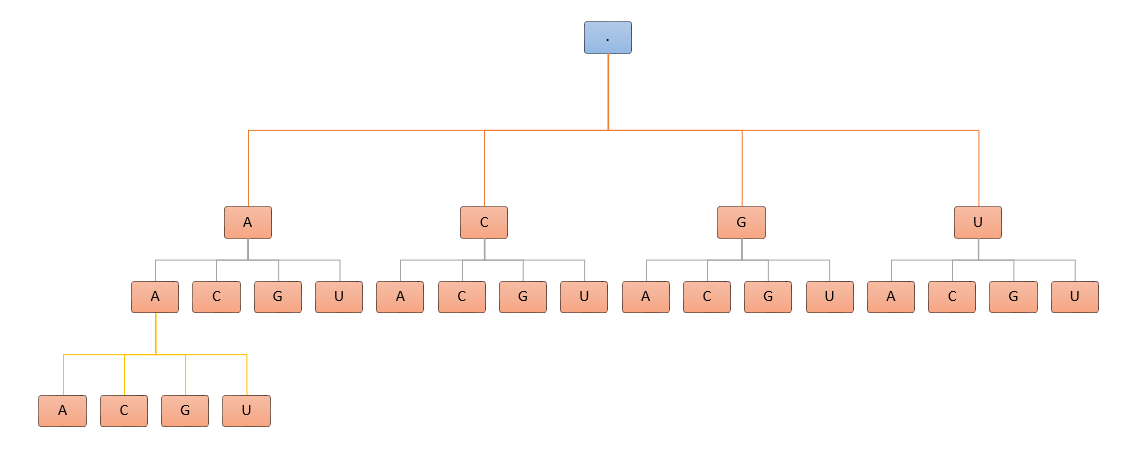}
    \caption[Example of a PCT of order 3 for $|E|=4$.]{Illustrative example of a PCT of order 3 for the dependency structure of nitrogenous bases, A C G T, in a string of RNA .}
    \label{fig:pct}
    \end{figure}
    
    Figure \ref{fig:pct} exemplifies a PCT of order 3 in an RNA example. It also shows that a tree representation offers easy interpretability of the dependency structure of a process. Clearly, a full Markov chain would require more parameters to accommodate the longer memory needed in one ``direction". In this example, only four contexts have a length of 3 while 15 other contexts have a length of 2, totaling 19 contexts. Completing the leaves for a full tree would result in a tree with 64 contexts. The PCT model is very beneficial from a data compression standpoint. Still, other applications in biology \cite{BEJERANO2001, BUSCH2009} and linguistics \cite{GALVES2012} have shown the value of this methodology to real-life applications.

    Besides the novel concept of only considering the relevant past, perhaps the biggest contribution of Rissanen's work was the proposal of the \textit{algorithm context} to estimate the true context tree given a finite sample. The true PCT, denoted by $\mathcal{T}_0$, contains the minimal set of strings needed to specify the probability of the next symbol completely. Several studies have built on this idea, either improving the results of the original paper \cite{BUHLMANN1999, DUARTE2006, GARIVIER2011}, or modifying the original algorithm \cite{WILLEMS1995, MARTIN2004}. Finding the true PCT through information criteria was thought to be computationally infeasible by \cite{BUHLMANN1999} because it would require comparing a very large number of hypothetical trees. The work of \cite{CSISZAR2006a} proves that it is indeed possible using the clever use of tree techniques. 
    
    Like the PCT model, the VNRF model also works with the idea of \textit{contexts}. Here, context is a minimal neighborhood needed to determine the probability of a site—the neighborhood's depth changes according to its values. Hence, the VNRF model is defined by a family of conditional probabilities that do not depend on a fixed neighborhood depth. In \cite{LOCHERBACH2011}, the focus was estimating the radius containing a site's minimal neighborhood. They do not address the problem of estimating the geometrical structure of the context, as they claim it would introduce too many parameters. Similarly, \cite{CSISZAR2006b} offers a consistent estimator for the context neighborhood of a site. Their paper, however, is mainly concerned with proposing a model selection criterion for MRFs since penalized likelihood estimators cannot be used. 
    
\subsection{Model Selection for Discrete MRFs}
\label{sec:pic}

    Analogous to the Bayesian Information Criterion (BIC), the pseudo-Bayesian information criterion (PIC) was proposed in \cite{CSISZAR2006b} to address the model selection problem in MRFs. The likelihood in BIC was replaced by the pseudo-likelihood introduced by \cite{BESAG1975}. Due to phase transition on multidimensional lattices, a unique invariant measure is not assured, so a likelihood approach is unsuitable. A  similar criterion was proposed earlier by \cite{JI1996}, and recently, \cite{PENSAR2017} introduced a small sample analytical version of PIC. The evaluation of the best model selection criteria for MRFs is beyond the scope of this work, and we will only focus on the definition and results related to PIC.
    
    \begin{definition}
    \label{def:pic_og}
    Let  $x(\Lambda_n)$ be a sample of an MRF such that 
    $\Lambda_1  \subset  \Lambda_2, ... ,\Lambda_n ,$
    $  n \in N$. 
    The pseudo-Bayesian information criterion (PIC) of a neighborhood $\Gamma$ is:
    \begin{align}
    \label{eq:pic_og}
        PIC_{\Gamma}\big(x(\Lambda_n)\big) = - \log MPL_{\Gamma} \big(x(\Lambda_n)\big) + |A|^{|\Gamma|}\log|\Lambda_n|
    \end{align}
    where $MPL_{\Gamma}$ is the maximum pseudo-likelihood, $\Lambda_n$ is the sample region,{ and $n$ is the number of sites in the sample}.
    \end{definition}

    \cite{CSISZAR2006b} proved that minimizing PIC over a family of hypothetical neighborhoods resulted in an estimate that equaled the true context neighborhood eventually almost surely as $n\to\infty$. The radius of the possible neighborhoods was allowed to grow with the sample size as $o((\log|\Lambda_n|)^{\frac{1}{2d}})$. This result is unaffected by phase transition and non-stationarity of the joint distribution. 
    
    The problem, however, is that no algorithm was proposed to compute the PIC estimator $\hat{\Gamma}_{PIC}$. This happened for two reasons. First, no simple formula is available for $|A|^{|\Gamma|}$ because the candidate neighborhoods do not have a specific geometry. The only requirement is that the neighborhood of a site $i$, denoted by $\Gamma_i$, is a finite central-symmetric set of sites with $i \notin \Gamma_i$. The second reason is that, even if it could be calculated, the authors did not find a way to compute the PIC score for all possible neighborhood configurations without calculating them one by one. Consequently, they leave it open if the PIC estimator can be computed in a ``clever way", as it was done in the one-dimensional case.

    That is precisely what the PCN model we propose here does for lattices in $\mathbb{Z}^2$. The PCN model is a two-dimensional version of a PCT that sets a fixed neighborhood geometry and represents the dependency structure as a tree. Consequently, the PCN algorithm is a modified version of the PCT algorithm in \cite{CSISZAR2006a}, using PIC instead of BIC to find the optimal tree.
    
    \section{Probabilistic Context Neighborhood  Model}
    \label{sec:pcn}

   The PCN model proposes a tree representation for MRF process on lattices in $\mathbb{Z}^2$, which is similar to the probabilistic context tree (PCT) model proposed by Rissanen (1983). The purpose of this model is to provide insight into the dependency of sites on their neighbors, through learning the dependency structure, as well as estimating the conditional probabilities that determine the value of a site. 
    
\subsection{Definitions and Notations}  
\label{sec:def}

      We consider a MRF in lattices in $\mathbb{Z}^d$ for the specific case where $d=2$. However, an important aspect of the PCN model is that the neighborhood geometry $\Gamma_i$ in Equation~\eqref{eq:mrf} is set to a \textit{frame}, denoted by $\partial_i^j$, as defined in \ref{def:frame}.
 
    \begin{definition}
    \label{def:frame}
    A frame $\partial^j_i$, with order $j \in \mathbb N$, is a particular type of neighborhood for a site $i$. It can be obtained by taking a square of side $ 2j+1 $, and removing a smaller square of side $ 2j-1 $ contained within it, both centered on $i$.
    \end{definition}
    
    \begin{figure}
        \centering
        \includegraphics[scale = 0.6]{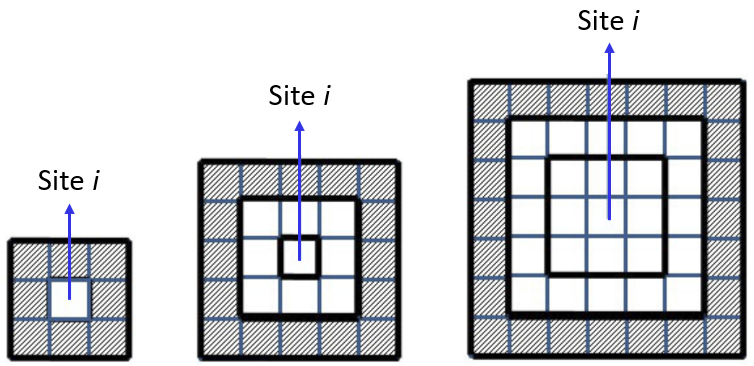}
        \caption{Frame structure $\partial^j_i$ for $j = 1, 2 $ and $3$, respectively.}
        \label{fig:frames}
    \end{figure}
    
    Figure \ref{fig:frames} provides an example of frames of order 1, 2, and 3. Larger orders can be understood analogously. It can be easily seen that, for $j=1,2,\dots,m$, the frames $\partial^j_i$ are nested sets. $\bigcap_{j=1}^m \partial^j_i= \emptyset$ and $\bigcup_{j=1}^m \partial^j_i$ is a square region of the lattice with side $2m+1$ and centered on site $i$. Since the geometry of the neighborhood is fixed and to simplify the notation, we will write $\partial^j$, omitting the site $i$ whenever it is clear. 
    
    We denote the union of frames $ \bigcup_{s=m}^n \partial^s = (\partial^m \partial^{m +1} \ldots \partial^n) $ as
    $\partial^ {m, \ldots, n}$, with $m<n$. The length of a frame is represented as $l(\partial^ {m, \ldots, n})=n-m+1$. For simplicity, the concatenation of the first frame with all the higher order frames until the $j$\textsuperscript{th} frame, given by $\partial^{1,\ldots,j}$, will be denoted as $\mathcal{D}^j$. The length of $\mathcal{D}^j$ is $l(\mathcal{D}^j)=j$ and equals the order of the neighborhood $\mathcal{D}^j$.
    
    We say that a configuration  $a(\partial_i^j)$ is a realization of the process on the subset $\partial_i^j$. The concatenation of two configurations $a( \partial^ {1, \ldots, k} )$ and $a(\partial^{m, \ldots, n}) $ is  $a(\partial^{1, \ldots, n})$, or $a(\mathcal D ^{ n}) $,  and is only possible if $m=k + 1$. The cardinality of a neighborhood, denoted by $|a(\mathcal{D}^n)|$, indicates the number of sites within a neighborhood of order $n$.
    
    \begin{definition}
    \label{def:suffix}
    A configuration $a(\mathcal D ^{ k}) $ is a \textit{suffix} of $a(\mathcal D ^{ n}), k \leq n $, if  $a(\mathcal D ^{ n})$ is a concatenation of $a( \partial^ {1, \ldots, k})$ and $a (\partial^{k+1, \ldots, n})$. This induces an order in the space of configurations and we say that  $ a(\mathcal D ^{ n}) \succeq  a(\mathcal D ^{ k})$. If the cardinality $|a (\partial^{k+1, \ldots, n})|> 0$, then  $a(\mathcal D ^{ k})$ is a \textit{proper suffix} of   $a(\mathcal D ^{ n})$. 
    \end{definition}
    
    A set of neighborhood configurations can be represented as a neighborhood tree $\mathcal{T}$. It has the root on top, characterizing the value of a site (identified as $\emptyset$), and branches connected to it, growing downwards. The first set of nodes stemming from the root is the first-order neighborhood configurations $\partial^1$. The \textit{children} of those nodes are the second-order neighborhood frames containing the \textit{parent} neighborhood frame inside, that is $\partial^{1,2}$ or simply $\mathcal{D}^2$. The third set of notes are the children of the second-order nodes, given by $\mathcal{D}^3$. The same logic is valid for higher-order nodes. 
    A neighborhood configuration $a(\mathcal{D}^j) \in \mathcal{T}$ represents a \textit{leaf} of the neighborhood tree. The leaves correspond to the last nodes of each of the branches connected to the root. Therefore, an internal node of $\mathcal{T}$ is a proper suffix of a leaf.
    
    As stated in Section \ref{sec:mrf}, all possible configurations of a random field $\{X(i),i\in\mathbb{Z}^2\}$, that take values in a finite alphabet $A$, are given by $\Omega = A^{\mathbb{Z}^2}$. Therefore, the number of possible neighborhood configurations of order $1$ in the PCN model is given by $A^{|\mathcal{D}^1|}$. The number of possible configurations of a neighborhood of order $2$ is $A^{|\mathcal{D}^2|}$ and so on. Hence, the formal definition of a neighborhood tree $\mathcal T$ is given below.
    
    \begin{definition}
    \label{def:tree}
    A subset $ \mathcal T  \subset \cup_{j=1}^\infty A^{|\mathcal D ^{ j}|} $ is called  a neighborhood tree if no  $ a(\mathcal D ^{ k}) \in \cal T $ is a suffix of any other $ a(\mathcal D ^{ n}) \in \mathcal T $.
    \end{definition}

    The depth of a neighborhood tree $\mathcal T$ represents the maximum order of neighborhoods belonging to that tree and is denoted by $d (\mathcal T) = \max_j \{ \, a(\mathcal D^{ j})\in \mathcal T\} $.
    
    
    If not a single neighborhood $a(\mathcal{D}^{j})$ belonging to the neighborhood tree $\mathcal T$ can be replaced by a proper suffix without violating the tree property, then the neighborhood tree is considered irreducible. The set of irreducible neighborhood trees is denoted by  $ \mathcal I $.
    
    
    Although the neighborhood geometry is fixed in a frame format, the order of the neighborhood needed to determine the probability of a site can still vary. Thus, the PCN model utilizes the VNRF framework and the notion of contexts as specified in Definition \ref{def:context}. 
    
    \begin{definition}
    \label{def:context}
    A finite configuration $ a(\mathcal D ^{j}) \in A^{|{\mathcal{D}}^{j}|}$ is a context neighborhood of a  Markov random field if $Q \big( a(\mathcal{D}^{j})\big) >  0$ and
    \begin{eqnarray}
    \label{eq:context}
    P \big(X(i) = a(i) \, \big| \, X (\mathbb Z^2 \backslash i) = a (\mathbb Z^2 \backslash i) \big) &=& P \big (X (i) = a (i) \, \big| \, X (\mathcal D ^{ j}) = a (\mathcal D^j) \big)\nonumber\\
    &=&Q\big(a(i)\, \big| \, a (\mathcal D ^{ j}) \big)
    \end{eqnarray}
    for every $ a(i) \in A $, and no proper suffix of $a(\mathcal D ^{j})$ has this property. 
    \end{definition}
    
    Therefore, if $a(\mathcal{D}^j)$ is a context neighborhood of a site $i$, then the probability distribution of that site depends only on $a(\mathcal{D}^j)$. There is no need to inspect the entire lattice to acquire information about the value assumed by $X(i)$. We say that $j $, which is the number of frames in the configuration $ a(\mathcal D ^{ j})$, is the \textit{order} of the context neighborhood. 
    
    
    
    Clearly, the set of all context neighborhoods of a process can be represented as a context neighborhood tree and we will denote it by $\mathcal{T}_0$. Let $Q_0 = \{\,Q(a(i)\,|\,a (\mathcal D ^{ j})): a(i)\in A, \, a (\mathcal D ^{ j}) \in \mathcal{T}_0  \}$ be the family of transition probabilities satisfying Equation \eqref{eq:context}. The pair $(\mathcal T_0, Q_0)$ is called \textit{probabilistic context neighborhood} or PCN.
    
    The goal of the PCN model is, given a finite sample $a(\Lambda_n)$ of a lattice in $\mathbb{Z}^2$, to estimate the PCN $(\mathcal T_0, Q_0)$ that generated the sample. In order to do so, the PIC score of \cite{CSISZAR2006b} is used to compare a set of hypothetical PCNs ($\mathcal{T}$, $Q$) to reach the true PCN $(\mathcal T_0, Q_0)$ that generated the sample under study.
    
    From now on, for simplicity, we refer to the PCN $(\mathcal T, Q)$ only as $\mathcal T$.
    
\subsection{Illustrating a PCN {$\mathcal{T}$}}
\label{sec:example}
    
    This section is dedicated to exemplifying the concepts and ideas defined in Section \ref{sec:def}. We focus on the space of binary states due to its simplicity and because it allows the interesting study of black-and-white images. An extension to larger state spaces is straightforward. 
    
    Let $A = \{ -1, 1 \}$, where $X(i) = -1$, if the value of site $i$ is white, and $X(i) = 1$ if it is black. 
    
    We consider two neighborhood configurations to be equivalent if each neighborhood contains the same number of black and white sites, independently of their position. 
    
    \begin{figure}
        \centering
        \subfloat[
        All possible configurations of first-order frames $a(\partial^1)$ for $A=\{-1,+1\}$.]{\label{fig:1frame-bw} \includegraphics[scale = 0.4]{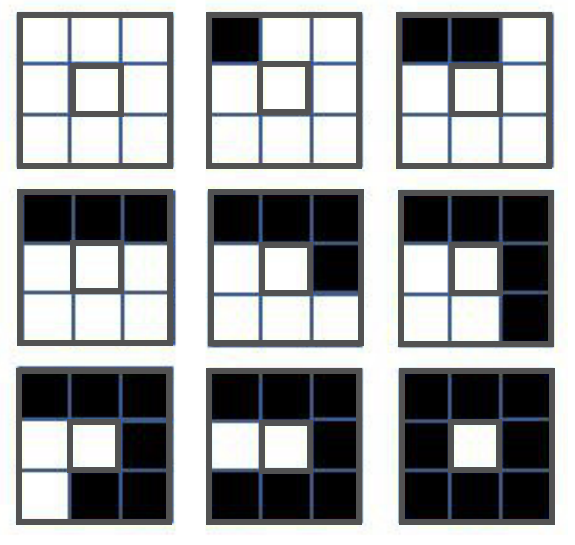} }%
        \qquad
        \subfloat[
        All possible configurations of second-order frames $a(\partial^2)$ for $A=\{-1,1\}$.]{\label{fig:2frames-bw} \includegraphics[scale = 0.3]{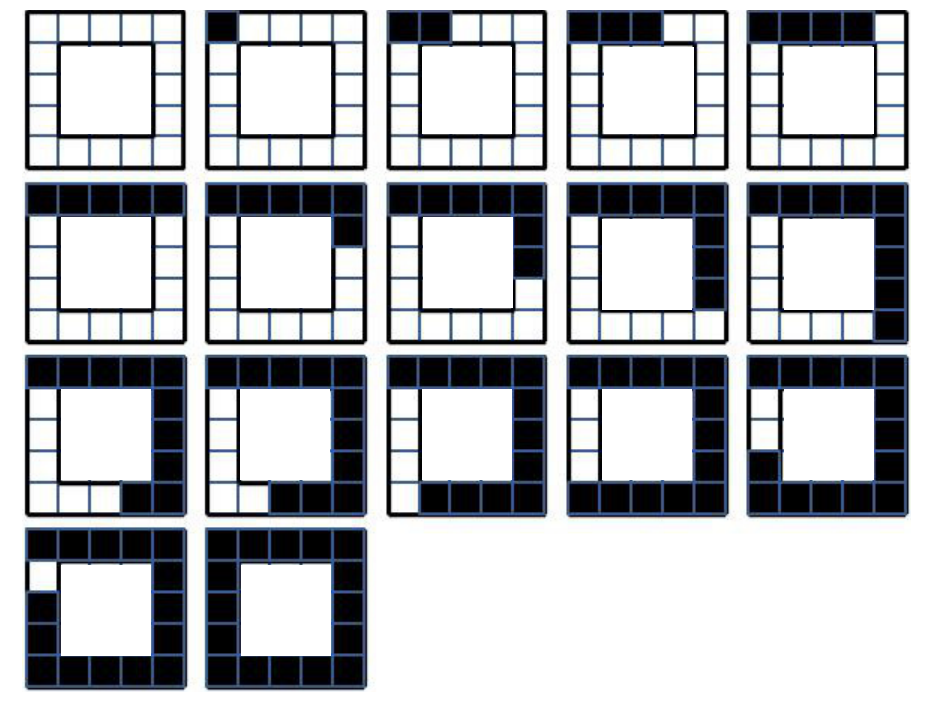} }%
        \caption{All possible configurations of first and second-order frames for black and white images.}%
        \label{fig:frames-bw}
    \end{figure}
    
     Figure \ref{fig:frames-bw} shows the possible neighborhood configurations for frames of order 1 and 2, respectively. It can be seen that a frame $\partial^1$ is made of 8 sites, that is, $|\partial^1| = 8$. Therefore, in the case of black and white images, there are 9 total possible configurations of first-order frames. The first frame can have zero black sites, all the way up to 8 black sites. In the case of frames $\partial^2$, there are 16 sites within it ($|\partial^2|=16$), which translates into 17 possible second-order frame configurations (varying from zero black sites all the way up to 16 black sites). Generalizing, the $j$\textsuperscript{th}-order frame has a total of $8j$ sites within it and $8j+1$ possible configurations.
    
\begin{figure}
        \centering
        \includegraphics[height=16em, width=42em]{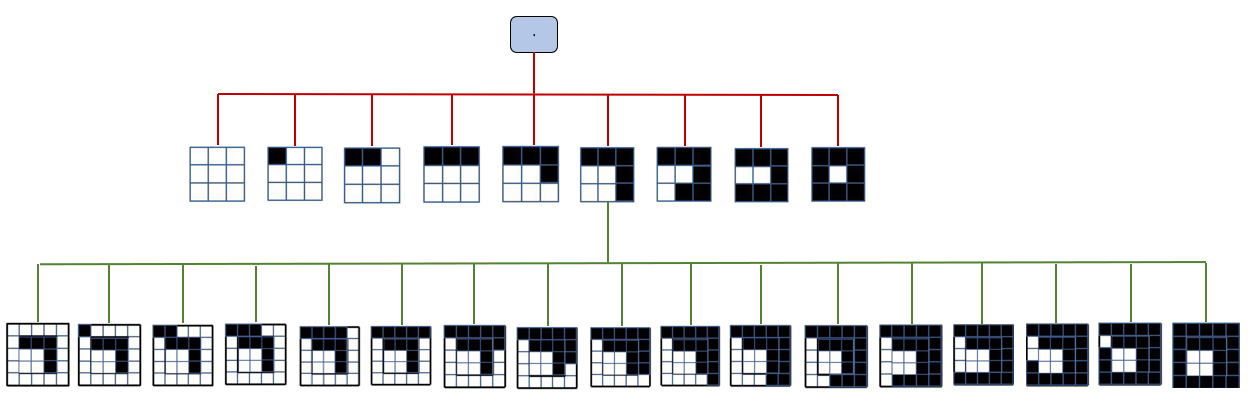}
        \caption{Illustrative example of a PCN $\mathcal{T}$ for $|A|=2$ and $d(\mathcal{T})=2$.}
        \label{fig:ex:pcn}
    \end{figure}

    The frame neighborhood geometry we propose makes it possible to represent the contexts of a MRF process in a tree format, similar to the PCT model. 
    A hypothetical PCN $\mathcal{T}$ for $A = \{ -1, 1 \}$ is shown in Figure \ref{fig:ex:pcn}. 
    
    The PCN root drawn on top of the tree represents the value of the site $i$. The first-generation nodes (children) are drawn from the root down and represent the first-order neighborhoods. If the information contained within the first-order frame is insufficient to provide a conditional probability for the site $i$, then the second-order neighborhood is drawn adding a frame of order 2 to this first-order neighborhood. The new neighborhood drawn is connected to the parent neighborhood. Each generation in the tree represents an added frame to the parent generation. The PCN tree continues to grow until all the context neighborhoods are added.
    
    In the example shown in Figure \ref{fig:ex:pcn}, the contexts of the PCN tree have variable neighborhood lengths. There are 8 contexts of order 1 and 17 contexts of order 2. For each context neighborhood, a conditional probability of the central site being black (or white) is assigned as in Definition \ref{def:context}. All first-order frames are considered contexts, except for the first frame with 5 black sites in it. This means that, if we observe only one black site in the first-order neighborhood (or 0, 2, 3, 4, 6, 7, and 8 black sites), it will be sufficient to determine the probability of the site $i$ being black. However, if there are 5 black sites in the first frame, we must continue ``down'' the PCN and look at the configurations of the second-order frame. All 17 child configurations of the first frame with 5 black sites are considered contexts. In summary, this hypothetical PCN $\mathcal{T}$ has depth $d(\mathcal{T})=2$, a total of  25 contexts neighborhoods (or leaves), and 1 internal node.

\subsection{Estimation in the PCN model}
\label{sec:PCNresults}

    We have explained and illustrated the neighborhood geometry and tree representation of an MRF process in the PCN model. This section will focus on the estimation procedure for a PCN $\mathcal{T}_0$ from a sample $a(\Lambda_n)$ containing the $n$ sites under study.
    
    As a likelihood approach is not suited for MRF  since we can not guarantee the existence of a unique invariant measure due to phase-transition problems, we use the pseudo-Bayesian information criterion of \cite{CSISZAR2006b} to select the optimal PCN $\mathcal{T}$. This is achieved by replacing the likelihood with the pseudo-likelihood introduced in \cite{BESAG1975}.

    \begin{definition}
    Given a sample $ a(\Lambda_n) $, the pseudo-likelihood function associated with a PCN $ (\mathcal {T}, Q) $ 
    is defined by:
     \begin{eqnarray}
     \label{eq:PL}
        PL_{\mathcal T}(a(\Lambda_n)) &=&  \prod_{a(\mathcal D^{j}) \in \mathcal T,  \, N_n(a(\mathcal D^{j}))\geq 1} \; \prod_{a(i)\in A} Q\big(a(i)\big|a(\mathcal D^{j})\big)^{N_n(a(\mathcal D^{j}, \,i))},\nonumber 
     \end{eqnarray}
    where
    $$N_n\big(a(\mathcal D^{j}, i)\big)=\big| \{ i \in a(\Lambda_n) :
    a( \mathcal D_i^{j}) \subset a(\Lambda_n), \, a(\mathcal D_i^{j}\cup {i})=a(\mathcal D_i^{j}, i)\}\big|$$
    represents the number of times that the configuration $a(\mathcal D^{j})$ is observed in the sample when the site $i $ assumes the value $a(i) $ and
    $$N_n\big(a(\mathcal D^{j})\big)=\big| \{ i \in a(\Lambda_n) : a(\mathcal D_i^{j}) \subset a(\Lambda_n)\}\big|$$
    is the number of occurrences of the configuration $ a(\mathcal D ^{ j}) $ in the sample $ a(\Lambda_n) $.
    \label{def:pseudolike}
    \end{definition}
    
    According to \cite{CSISZAR2006b}, the maximum pseudo-likelihood is obtained for:
    \begin{align}
    \hat{Q}\left(a(i)\big|a(\mathcal{D}^j)\right) = \frac{N_n\left(a(\mathcal{D}^j,i)\right)}{N_n\left(a(\mathcal{D}^j)\right)} \nonumber
    \end{align}
    
    Therefore, given a sample $a(\Lambda_n)$, the maximum pseudo-likelihood ({MPL}) for a PCN $\mathcal T$ is:
    \begin{align}
    \label{eq:MPL}
    MPL_{\mathcal T}\big(a(\Lambda_n)\big) = \prod_{a(\mathcal{D}^{j}) \in \mathcal T,  \, N_n(a(\mathcal D^{j}))\geq 1} \; \prod_{a(i)\in A} \left( \frac{N_n\left(a(\mathcal{D}^j,i)\right)}{N_n\left(a(\mathcal{D}^j)\right)}\right)^{N_n\left(a(\mathcal{D}^j,i)\right)}
    \end{align}
    
   Since we are interested in estimating the PCN $\mathcal{T}_0$, instead of the neighborhood $\Gamma$, we modified the PIC formula in Equation~\eqref{eq:pic_og} to be closer to the BIC formula for PCTs, replacing the maximum likelihood by the maximum pseudo-likelihood. 
    
    \begin{definition}
    \label{def:pic}
    Given a sample $ a (\Lambda_n) $, the pseudo-Bayesian information criterion (PIC) for a PCN $ \mathcal T $ is:
    \begin{align}
    \label{eq:pic}
    PIC_{\mathcal T}\left(a(\Lambda_n)\right)=-\log MPL_{\mathcal T}\left(a(\Lambda_n)\right) + \frac{(|A|-1)|\mathcal T|}{2} \log|\Lambda_n|
    \end{align}
    \end{definition}

    An important difference between the definition above and Definition \ref{def:pic_og} is the term that precedes $\log|\Lambda_n|$. Because the neighborhood structure in \cite{CSISZAR2006b} was not fixed, it was unfeasible to compute the term $|A|^{|\Gamma|}$. In the PCN model, however, the fixed frame geometry for the neighborhoods allows the computation of $|\mathcal T|$, which represents the number of leaves of a PCN tree or simply the number of neighborhood contexts $a(\mathcal D^j) \in \mathcal{T}$.  Our work obtained a closed formula for $|\mathcal{T}|$.

When the position of the symbol matters, each site can receive $|A|$ symbols, as there are $8^k$ sites, so we will have a total of $|A|^{8k}$ distinct configurations for each order $k$. For example, if $k=1$ and $A=\{ 0,1\}$, we  have $|\mathcal{T}|= 256$.


We can significantly reduce the number of possible configurations by considering that only the number of symbols in a frame matters, not their position. To do this, we need to count the ways in which we can distribute $8^k$ sites into $|A|$ groups. For the case where $k=1$ and $A=\{0,1\}$, we need to divide the $8$ sites in the order one frame into two groups. To do this, we can place a separator between the numbers $1$ to $8$. The placement of the separator corresponds to the number of $"0"$ symbols in the configuration. For instance, placing the separator after the $1$ implies that the configuration has one $"0"$ symbol and seven $"1"$ symbols. The position of the separator indicates the number of $"0"$ symbols in the configuration. This way, we have nine positions (eight sites and one separator), and we want to count how many ways we can distribute this separator among the eight sites, which gives us $\binom{8+1}{1} =9$ configurations. Similarly, if we have $|A|$ symbols and order $k$, we will have $\binom{8^k + |A|-1} {|A|-1}$ possible configurations for each $k$. Therefore, if frames are considered equivalent by having the same combination of elements in $A$ within a frame, then $|\mathcal {T}|$ is, at most,
    \begin{align}
    \label{eq:leaves}
    \prod_{k=1}^{d(\mathcal T)}  \binom{{8k+|A|-1}}{{|A|-1}}
    \end{align}


    Since Definition \ref{def:frame} states that the k-{th} frame is obtained by taking a square of side $2k+1$ and removing a smaller square of side $2k-1$, both centered on site $i$. Therefore, a neighborhood $\mathcal{D}^k$, which is the concatenation of frames of order 1 through k, is given by the number of sites within a square of side $2k+1$ minus the center site:
    \begin{align}
    |\mathcal{D}^k| = (2k+1)(2k+1) - 1 = 4k^2 + 4k = \frac{8k(k+1)}{2}. \nonumber
    \end{align}
    
    Therefore, the number of leaves of a PCN $\mathcal{T}$ of depth $d(\mathcal{T}) = k$ is the number of possible arrangements of $|\mathcal{D}^k|$ sites, where each site can assume $|A|$ possible values.
    \begin{align}
    |\mathcal{T}| = |A|^{|\mathcal{D}^k|} =|A|^{\frac{8k(k+1)}{2}}. \nonumber
    \end{align}

    Conversely, if we consider that the position of each site within the frame matters, then $|\mathcal{T}|$ is, at most, 
    \begin{align}
    |A|^{|\mathcal{D}^k|} =|A|^{\frac{8k(k+1)}{2}} \quad \text{, where}\, k=d(\mathcal{T}). \nonumber
    \end{align} 
    
    Consequently, the PCN model solves the first issue of computing the PIC score for a given PCN $\mathcal{T}$ since we can calculate the penalizing term in Equation (\ref{eq:pic}).
    
   Given a sample $ a (\Lambda_n) $, a feasible PCN $\mathcal{T} $ is such that $ d(\mathcal{T} )\leq D (n) $, where $D(n)$ is an appropriate function of the sample size. Also, for every $ a(\mathcal{D}^{ j}) \in \mathcal{T} $, $N_n(a (\mathcal{D}^{ j})) \geq 1$. We say that $ a(\mathcal D ^k) $ is a suffix of some $a(\mathcal{D}^{ j}) \in \mathcal{T} $ if $k\leq j$ and $N_n (a (\mathcal{D}^k)) \geq 1$. The family of feasible PCNs is denoted by $ \mathcal{F}_1 \left(a (\Lambda_n), D (n)\right) $.
    
    \begin{definition}
   \label{def:Tpic}
    We define the PIC estimator for a PCN $\mathcal{T}_0$ as
        \begin{align}
       \label{eq:Tpic}
        \hat {\mathcal{T}}_{PIC}\left(a(\Lambda_n)\right) =  \argmin_{\mathcal{T} \in \mathcal{F}_1
        \left(a(\Lambda_n),D(n)\right) \cap \mathcal I} PIC_{\mathcal{T}} \left(a(\Lambda_n)\right),
       \end{align}
    \end{definition}

    In other words, the PIC estimator for a PCN $\mathcal{T}_0$ is the PCN $\mathcal{T}$ that minimizes the PIC score among all feasible PCNs allowed to grow with the sample size. 

In the case where the PCN has only one invariant measure, the consistency of the $\hat {\mathcal{T} }_{PIC}$ estimator is guaranteed in a similar way to that of the $\hat {\mathcal {T}}_{BIC}$ estimator for the case of one-dimensional PCTs (See  Theorem 2.6 in \cite{CSISZAR2006a}).
    

    Suppose we cannot guarantee the existence of a single invariant measure for the PCN. In that case, we still have the result proven in \cite{CSISZAR2006b}, Theorem 2.1, that guarantees the consistency of the neighborhood estimated using the PIC ($\hat{\Gamma}_{PIC}$). This way, the tree  $\hat{\mathcal{T} }_{PIC}(a(\Lambda_n))$ is a finite union of consistently estimated neighborhoods $\Gamma_i$ of site $i$ (frames).
    
     The authors also prove that the empirical estimator $\hat Q\left(a(i)\,\big|\, a(\hat\Gamma_{PIC})\right)$ converges to the true conditional probability almost surely as $n \to \infty$. However, the question of {\it how to find the PIC estimator without computing the score for all possibilities?} was unanswered.

\subsection{PCN algorithm}
\label{sec:PCNalgorithm}

   Calculating PIC for all feasible PCNs $\mathcal T$ would be impractical and time-consuming. We propose a PCN algorithm analogous to the one initially proposed by \cite{CSISZAR2006a} for the one-dimensional case in a clever way. The PCN model represents the context neighborhoods of an MRF in a tree format, similar to the PCT model. Then we adapted the PCT algorithm to an MRF framework, which made it possible to obtain $\hat{\mathcal{T}}_{PIC}$. Our main contribution is presenting a way to convert the two-dimensional neighborhood dependence in MRF in lattices into a tree format, depending only on the size order $j$, which allows us to borrow the clever ideas presented in \cite{CSISZAR2006a}.

 First, we need to express $\hat {\mathcal T}_{PIC}\left(a(\Lambda_n)\right)$  in Equation (\ref{def:Tpic}) in terms of the pseudolikelihoods $P_{MPL,\mathcal D^{j}}(a(\Lambda_n))$. To simplify notation we denote the set of all trees considered by  $\mathcal F = \mathcal F_1\left(a(\Lambda_n),D(n)\right) \cap \mathcal I$. Then, for a sample  $a(\Lambda_n)$, and using Definitions \ref{def:Tpic} and \ref{def:pic}, we have that:
    \begin{eqnarray}
    \hat {\mathcal T}_{PIC}\left(a(\Lambda_n)\right) 
    &=& \argmin_{\mathcal T\in \mathcal F} PIC_\mathcal T \left(a(\Lambda_n)\right) \nonumber\\
    &=& \argmax_{\mathcal T\in \mathcal F} \left\{ \, \log MPL_{\mathcal{T}}\left(a(\Lambda_n)\right) - \frac{(|A|-1)|\mathcal{T}|}{2} \log n \, \right\} \nonumber\\
    &=&  \argmax_{\mathcal T\in \mathcal F} \left\{ \,  n^{-\frac{(|A|-1)|\mathcal{T}|}{2}}\, MPL_{\mathcal{T}}\left(a(\Lambda_n)\right) \, \right\} \nonumber
    \end{eqnarray}
    
    The maximum pseudo-likelihood function in Equation~\eqref{eq:MPL} can be factorized as:
    \begin{align}
    \label{eq:MPL-fac}
    MPL_{\mathcal T}(a(\Lambda_n))= \prod_{a(\mathcal D^{j}) \in \mathcal T } \tilde P_{MPL,\mathcal D^{j}}(a(\Lambda_n)), \nonumber
    \end{align}
    where
    \begin{align}
    \tilde P_{MPL,\mathcal D^{j}}(a(\Lambda_n))=\left\{\begin{array}{lc}
    \displaystyle \prod_{a(i) \in A} \left(\frac {N_n\left(a(\mathcal D^{j},i)\right)}{N_n(a(\mathcal D^{j}))}\right)^{N_n(a(\mathcal D^{j},i))},&\mbox{if}\quad N_n(a(\mathcal D^{j})) \geq 1\\
    1 \qquad \qquad \qquad \qquad \qquad \qquad \quad, &\mbox{if}\quad N_n(a(\mathcal D^{j})) = 0
    \end{array}\right.\nonumber
    \end{align}
    
    Hence, the PIC estimator $\hat{\mathcal{T}}_{PIC}$ can be rewritten as:
    \begin{eqnarray}
    \label{eq:Tpic-alg}
    \hat {\mathcal T}_{PIC}\left(a(\Lambda_n)\right) 
     &=&  \argmax_{\mathcal T\in \mathcal F} \left\{ \,  n^{-\frac{(|A|-1)|\mathcal{T}|}{2}}\, \prod_{a(\mathcal D^{j}) \in \mathcal T} \tilde P_{MPL,\mathcal D^{j}}(a(\Lambda_n)) \, \right\} \nonumber \\
     &=& \argmax_{\mathcal T\in \mathcal F} \left\{ \, \prod_{a(\mathcal D^{j}) \in \mathcal T} \tilde P_{\mathcal D^{j}}\left(a(\Lambda_n)\right) \, \right\},
    \end{eqnarray}
    where $\tilde P_{\mathcal{D}^{j}} \left(a(\Lambda_n)\right) = n^{-\frac{|A|-1}{2}}\, \tilde P_{MPL,\mathcal D^{j}}(a(\Lambda_n))$.\\

 Following \cite{CSISZAR2006a}, we define certain auxiliary variables required for implementing the PCN algorithm. After each definition, we provide an example illustrating how to compute each variable.
    \begin{definition}
    \label{def:VdjXdj}
    Given a sample $ a(\Lambda_n) $, each neighborhood $a(\mathcal{D}^j) \in \mathcal{N}_j^D$ receives recursively, from the leaves of the tree to the root, the value
        \begin{eqnarray}
        \label{eq:vdj}
        V_{\mathcal D^{j}}^{D}(a(\Lambda_n))=\left\{\begin{array}{lr}
        
        \tilde P_{\mathcal D^{j}}(a(\Lambda_n))\qquad \qquad \qquad \qquad \qquad \qquad \qquad \qquad \qquad\ , \quad \mbox{if}\quad   j= D\\
        
        \max\left\{ \tilde P_{\mathcal D^{j}}(a(\Lambda_n)) \ ,\ \displaystyle \prod_{a(\mathcal{D}^{j+1}):\; N_n\left(a(\mathcal D^{j+1})\right)\geq 1   }V_{\mathcal D^{j+1}}^{D}\left(a(\Lambda_n)\right)\right\}, \quad  \mbox{if} \, \ 0\leq j< D
        \end{array}\right. \nonumber
        \end{eqnarray}
\end{definition}        
This variable compares the pseudo-likelihood values between parents and their children and returns the highest ones.

\begin{figure}[H]
 \centering
    \includegraphics[scale=0.25]{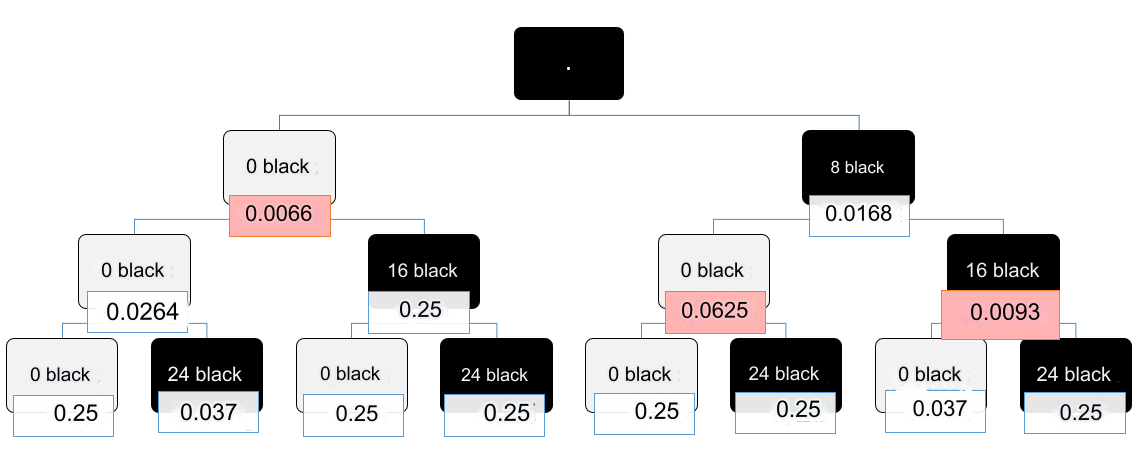}
    \caption{Illustration of the step of calculating $V_{\mathcal D^{j}}^{D}(a(\Lambda_n))$ for each neighborhood $a(\mathcal{D}^j) \in \mathcal{N}_D$.}
     \label{ex:tree}
\end{figure}

       To illustrate this step of the PCN algorithm, Figure \ref{ex:tree} presents an example of a tree with values of $V_{\mathcal D^{j}}^{D}$  computed for each node for a given sample $(a(\Lambda_n))$ where $D=3$. We color the nodes pink, where the product of the values of the children has higher values than their parents.

    \noindent The indicator $\chi_{\mathcal{D}^{j}}^D$ is associated with each node based on the values of $V_{\mathcal{D}^{j}}^{D}$, defined in \ref{def:VdjXdj}, to compare parent and child nodes in the following way.

    \begin{definition} For each node in $\mathcal{N}^D$, define

        \begin{eqnarray}
        \label{eq:xdj}
        \chi_{\mathcal D^{j}}^D(a(\Lambda_n))=\left\{\begin{array}{rl}
        
        0 ,&\mbox{if} \quad  j=D\\
        
        0 ,&\mbox{if} \quad      \tilde P_{\mathcal D^{j}}(a(\Lambda_n)) \geq \displaystyle \prod_{a(\mathcal{D}^{j+1}):\; N_n\left(a(\mathcal D^{j+1})\right)\geq 1   }V_{\mathcal D^{j+1}}^{D}\left(a(\Lambda_n)\right) \,\ \mbox{and} \, \ 0\leq j< D\\
        
        1  ,&\mbox{if}\quad  \tilde P_{\mathcal D^{j}}(a(\Lambda_n)) <  \displaystyle\prod_{a(\mathcal{D}^{j+1}):\; N_n\left(a(\mathcal D^{j+1})\right)\geq 1   }V_{\mathcal D^{j+1}}^{D}\left(a(\Lambda_n)\right) \,\ \mbox{and}\, \ 0\leq j< D \nonumber
        
        \end{array}\right.
        \end{eqnarray}
    where $a(\mathcal{D}^{j+1})$ represents the children of the parent neighborhood $a(\mathcal{D}^{j})$.
   
    \end{definition}

\begin{figure}[H]
    \centering
    \includegraphics[scale=0.25]{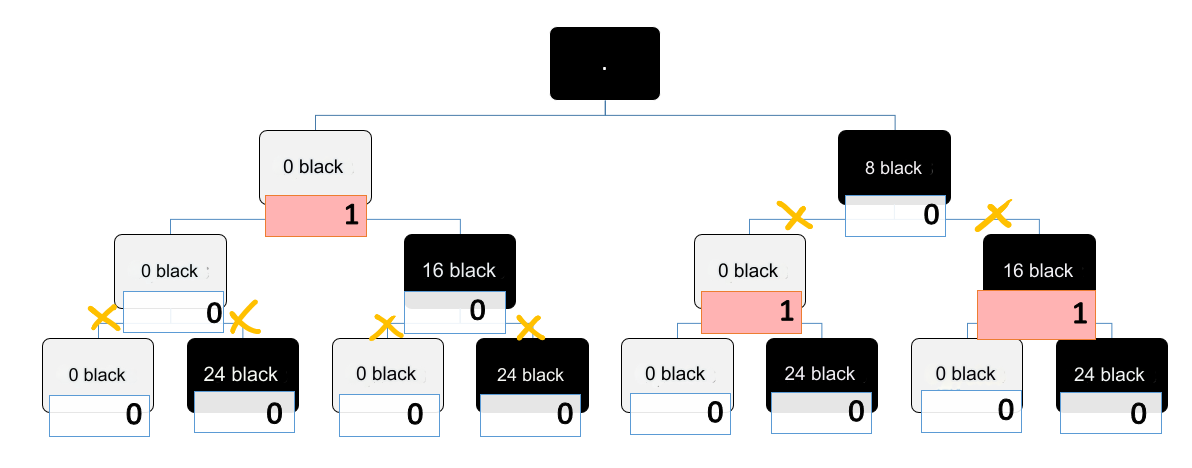}
    \caption{Illustrating tree pruning.}
    \label{ex:poda}
\end{figure}

    In Figure \ref{ex:poda}, we can see the indicators $\chi_{\mathcal D^{j}}^D$ that are associated with the nodes of the PCN shown in Figure \ref{ex:tree}. 
    
 Based on the indicators $\chi_{\mathcal D^{j}}^D(a(\Lambda_n))$, a maximizing tree $\mathcal{T}_{\mathcal{D}^j}^D\left(a(\Lambda_n)\right)$ comprised of neighborhoods $a(\mathcal{D}^u)\succeq a(\mathcal{D}^j)$ is defined in the  following way.

    \begin{definition}
    \label{def:Tdj}
    Given $a(\mathcal{D}^j) \in \mathcal{N}_j^D$, let $\mathcal{T}_{\mathcal D^{j}}^D\left(a(\Lambda_n)\right)$ equal to 
        \begin{eqnarray}
        \left\{\begin{array}{rl}
        a(\mathcal{D}^j) \qquad \qquad \qquad \qquad \qquad \qquad \qquad \qquad \qquad \qquad \qquad \qquad \qquad \quad,&\mbox{if} \;  \chi_{\mathcal D^{j}}^D(a(\Lambda_n))=0\\
        
        \left\{a(\mathcal{D}^u) \in \mathcal{N}_j^D\: :\, \chi_{\mathcal D^{u}}^D(a(\Lambda_n))=0,\; \chi_{\mathcal D^{v}}^D(a(\Lambda_n)) =1 , \text{ for all}\; j\leq v<u\right\} \: ,&\mbox{if}\; \chi_{\mathcal D^{j}}^D(a(\Lambda_n))=1\nonumber
        \end{array}\right.
        \end{eqnarray}
    \end{definition}

 According to  Definition \ref{def:Tdj}, we obtain the pruned tree $\mathcal{T}_{\mathcal D^{j}}^D\left(a(\Lambda_n)\right)$ by starting from the top and removing all bunches of children less significant than their parents at once after the first "0" (since "0" indicates that the children's pseudolikelihood product is smaller than their father's likelihood). This way  the pruning procedure becomes more efficient.

    

    Borrowing the ideas presented in \cite{CSISZAR2006a} for one-dimensional PCT, the probabilistic context neighborhood tree estimator $\hat {\mathcal T}_{PIC}(a(\Lambda_n))$ equals the maximizing tree obtained when the pruning procedure starts from the root ($ D^{j} = \emptyset$). That is,
    $$\hat {\mathcal T} _{PIC}\left(a(\Lambda_n)\right)=\mathcal{T}_{\emptyset}^D\left(a(\Lambda_n)\right)$$

In other words, the maximizing tree assigned to the root $\mathcal{T}_{\emptyset}^D(a(\Lambda_n))$ is the tree, among all the feasible trees, that maximizes the product $\prod_{a(\mathcal D^{j}) \in \mathcal T} \tilde P_{\mathcal D^{j}}\left(a(\Lambda_n)\right)$ in Equation~\eqref{eq:Tpic-alg}. 

\begin{figure}[H]
    \centering
    \includegraphics[scale=0.35]{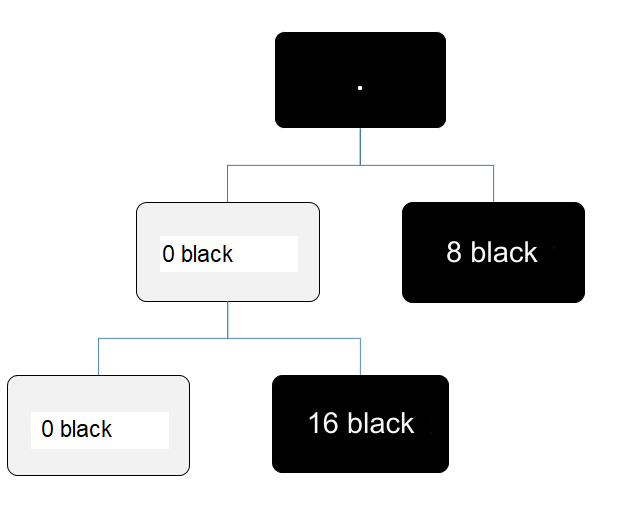}
    \caption{Illustrating the final result of the PCN algorithm.}
    \label{ex:finaltree}
\end{figure}

     Figure \ref{ex:finaltree} shows the final estimated tree after the pruning procedure.

    In short, the maximizing tree assigned to the root (or equivalently, the PIC estimator for PCN $\mathcal{T}_0$) can be obtained by pruning the tree containing all configurations that belong to the sample $a(\Lambda_n)$, $\mathcal{T}^D_{max}a(\Lambda_n)$,  as determined by Definition \ref{def:Tdj}. Unlike the assignment of values $V_{\mathcal D^{j}}^{D}(a(\Lambda_n))$ and indicators  $\chi_{\mathcal D^{j}}^D(a(\Lambda_n))$, the pruning procedure is done starting from the root of the tree and moving "down" the branches. The indicator $\chi_{\mathcal D^{j}}^D(a(\Lambda_n))$ determines where to prune the tree. If an indicator equals zero, we keep that specific node and exclude the children configurations connected to it. Alternatively, if the indicator of a node equals one, we continue "down" to the children's configurations until we observe an indicator equal to zero. That procedure is executed for all the branches connected to the root. So, after the pruning procedure is finalized, the resulting tree has internal nodes with indicators equal to one, and all the leaves have indicators equal to zero. \\

 \bigskip

   The steps for the PCN algorithm are then presented in { \bf Algorithm \ref{alg:pcn}}. 
   
    \begin{algorithm}[H]
    \caption{PCN algorithm}
        \label{alg:pcn}
        \textbf{ For a sample} $a(\Lambda_n)$ get $\mathcal{T}^D_{max}a(\Lambda_n)$ with $D=D(n)$;
        
        \textbf{ Define} $\mathcal{N}_j^D$ as the set of nodes of  $\mathcal{T}^{D}_{max}a(\Lambda_n)$ of order $j$;
        
        \textbf{ Compute} $\tilde P_{\mathcal{D}^{j}} \left(a(\Lambda_n)\right)$  for each $a(\mathcal{D}^j) \in \mathcal{N}_D, 1 \leq j \leq D $.

        \textbf{ Compute} $V_{\mathcal D^{j}}^{D}(a(\Lambda_n))$, $ 1 \leq j \leq D  $;
        
        \textbf{Compute} $\chi_{\mathcal D^{j}}^D(a(\Lambda_n))$ recursively for each $j$  from $ D $ to $1  $;
        
        \textbf{ Do} $\mathcal{T}^{D}_{max}a(\Lambda_n)= \mathcal{T}^{D}_{\emptyset}\left(a(\Lambda_n)\right)$;

\begin{algorithmic}[1]
\State $j \gets 1$
  \State For each node in $\mathcal{N}_j^D$  
\If{ $\chi_{\mathcal D^{j}}^D(a(\Lambda_n)) =0$} 
    \State  keep $a(\mathcal{D}^j) \;in \; {\tau^{D}_{\emptyset} \left(a(\Lambda_n)\right)}$ and prune   $a(\mathcal{D}^r)$ for $  r > j$
\Else
    \State{$j \gets j+1$} 
   \State{ Go to 3}
\EndIf 
\end{algorithmic}

  \textbf{ Return $\hat{\mathcal{T}}_{PIC} \left(a(\Lambda_n)\right) = \tau^{D}_{\emptyset} \left(a(\Lambda_n)\right)$.}
   
 \end{algorithm}
    
\subsection{ Building confidence intervals for the conditional probabilities in $\mathcal{T}_0$}
\label{bootIC}

 We apply a bootstrap technique to generate interval estimates for the conditional probabilities of an unknown PCN process $\mathcal{T}_0$. This method involves resampling from the estimated PCN $\hat{\mathcal{T}}$ obtained via the PCN algorithm. We generate larger samples from $\hat{\mathcal{T}} $ ( side $ n + \delta$) than the original sample ( side $n$ ) that produced the estimated tree and used this extra neighborhood when estimating the conditional probabilities of the tree's contexts in the boundaries. This ensures equal probability law to all sites, regardless of location, within or outside the boundaries.  After estimating the conditional probabilities of each context for all samples, we order them and exclude the smallest and largest $2.5\%$ values to obtain a   $95\%$ confidence interval. To ensure the correctness of our bootstrap estimation, we need to make sure that the value of $\delta$ is greater than the highest order of $\hat{\mathcal{T}}$, $d(\hat{\mathcal{T}})$. We present a general pseudocode of the bootstrap  confidence interval estimation using the PCN in Algorithm  \ref{boot}
 
 \begin{algorithm}
\caption{ Bootstrap Confidence intervals for $\mathcal{T}_0$ }
  \label{boot}
  \begin{algorithmic}[1]
      \Require{ $\delta > d(\hat{\mathcal{T}})$}
     \State{Generate B samples of size $(n + \delta) \times (n + \delta)  $ }
    \State{ While{ $i < B$}}
     \State{ Get $\hat{\mathcal{T}}^i$ for each sample $i$  }
    \State{ Extract the conditional probabilities of all contexts in  $\hat{\mathcal{T}}^i$}
    \State{Order the  conditional probabilities' values for each context of $\hat{\mathcal{T}}^i$}
   \State{Print the percentiles $0,025$ and $0,9725$}
   \end{algorithmic}
 \end{algorithm}

\section{Simulation Study}
\label{sec:sim}
    
    This section aims to validate the PCN algorithm explained in Section \ref{sec:PCNalgorithm}. We conducted a simulation study for two different scenarios using the statistical software R \cite{R}. We seek to compare the estimated trees obtained through the PCN algorithm with the original trees that generated the sample. 
    
    Our simulations are based on a regular lattice with black-and-white sites. We borrow the notation used in Section \ref{sec:example}, considering $A=\{-1,1\}$ where $a(i) = -1$, if the observed value of site $i$ is white, and $a(i) = 1$ if it is black. Since $|A|=2$, we have complementary events, and determining the conditional probability of a site being black suffices to determine the conditional probability of it being white. In addition, we also consider frames to be equivalent if they have the same number of black sites within them, just as in the example provided in Section \ref{sec:example}. 
    
\subsection{Generating samples}
\label{sec:mcmc}

    \begin{figure}[h]
        \centering
        \includegraphics[width=41em,height=16em]{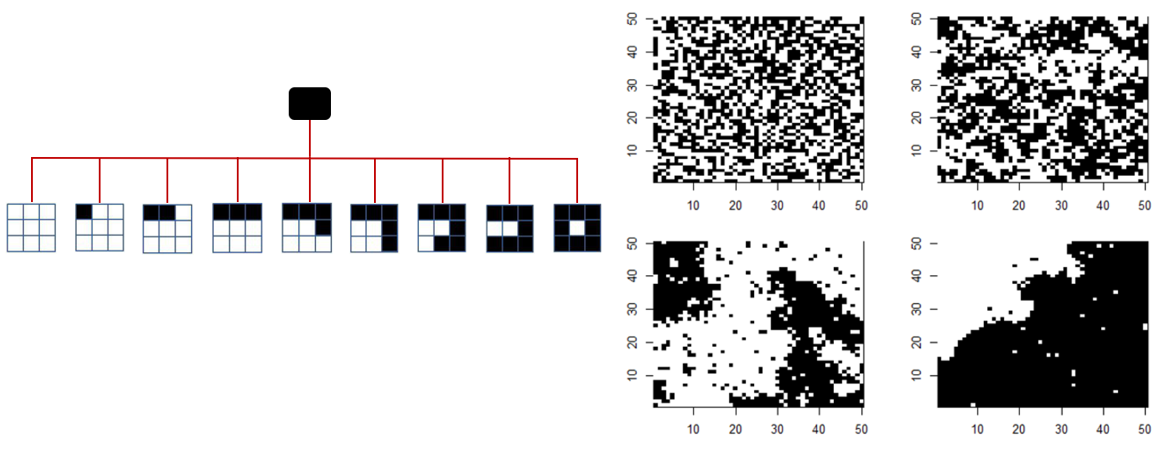}
    \caption{Left: Probabilistic context neighborhood tree structure. Right: Lattice simulations generated from the PCN structure on the left. For each black-and-white image shown on the right side, the tree structure was the same, the only variation was in the conditional probabilities assigned to the leaves.}
    \label{fig:samples}
    \end{figure}
    
    In order to generate samples with a predefined spatial dependency, we first determined the PCN $\mathcal{T}_0$'s structure and the conditional probabilities associated with each leaf. The same PCN tree structure can create different images when the conditional probabilities of each context neighborhood differ, as shown in Figure \ref{fig:samples}.
    
    Sampling is done using a Markov chain Monte Carlo ({MCMC}) method. Starting from a random configuration of black-and-white sites, we evaluate each site individually. A conditional probability of being black is attributed to a site based on its neighbors, as dictated by the PCN tree $\mathcal{T}_0$. An acceptance step, similar to the Metropolis-Hastings algorithm \cite{METROPOLIS1953,HASTINGS1970}, is then used to determine whether the site under evaluation is black or white. Once this procedure is done for all sites, we have completed the first iteration. We perform iterations until the image ``stabilizes". From that point on, we consider that the process has converged to the target distribution.
    
    In this study, the sites were inspected one column at a time, line by line. We conducted a random inspection of the lattice and found no significant difference in the total computational time or time until convergence.
    To ensure that the sites located on the boundaries are evaluated correctly, we mirrored the lattice both horizontally and vertically. In doing so, we guarantee that the sites outside the boundaries follow the same law as those inside the boundaries.

\subsection{Estimating a PCN {$\mathcal{T}_0$}}
\label{sec:estimation}
    
    This section presents the estimated PCN trees obtained through the PCN algorithm. For each scenario, a black and white image was simulated from a given PCN $\mathcal{T}_0$ as described in the previous section.

\subsubsection{Simulation 1: Variable-neighborhood PCN {$\mathcal{T}_0$} with {$d(\mathcal{T}_0)=2$} }
\label{subsec:sim2}

\begin{figure}[h]
	   \centering
	   \includegraphics[width=27em, height=18em]{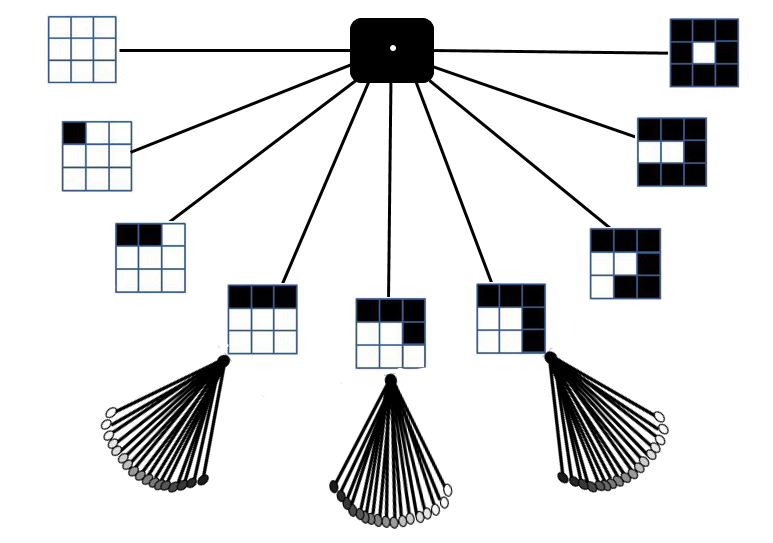}
	   \caption{Variable-neighborhood PCN $\mathcal{T}_0$ with depth $d(\mathcal{T}_0)=2$.}
	   \label{fig:sim2true}
    \end{figure}

 Our first simulation is based on a variable-neighborhood PCN $\mathcal{T}_0$ with $d(\mathcal{T}_0) = 2$. The source PCN $\mathcal{T}_0$ is shown in Figure \ref{fig:sim2true}. It has 6 first-order contexts and 51 second-order contexts neighborhoods. Each internal node of this tree has 17 children, representing all possible second-order frame configurations (that vary from 0 to 16 black sites within it). This PCN tree indicates that, if there are 3 black sites in the first frame (or 4 and 5), it is necessary to look at the second-order frame configuration to determine the transition probability for the given site. Due to space limitations, we choose not to draw the second-order configurations and draw a grayscale instead. The lighter the color, the less black sites exist in the second frame. On the other hand, the darker the color, the more black sites. 
    
    \begin{figure}[h]%
    \centering
    \subfloat[
    Sample lattice from a variable neighborhood PCN $\mathcal{T}_0$.]{\label{fig:sim2sample}\includegraphics[width=17em, height=17em]{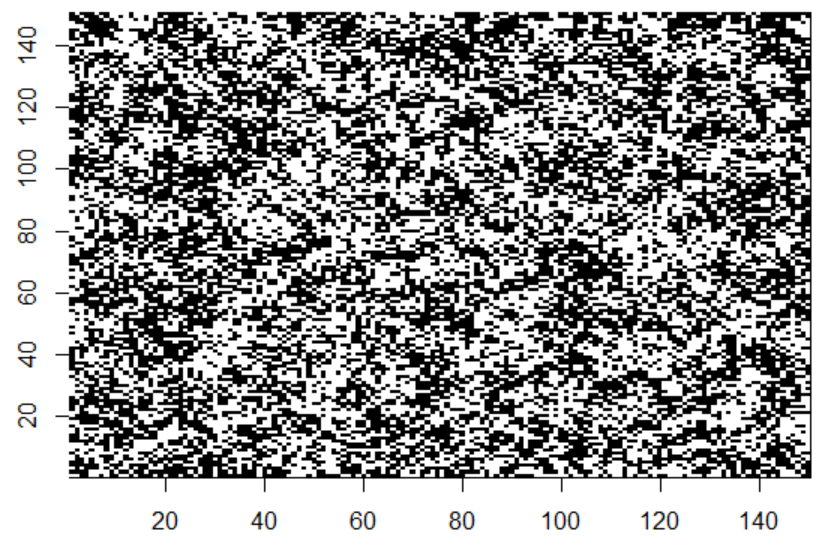} }%
    \qquad
    \subfloat[
    Estimated PCN $\hat{\mathcal{T}}$ after running the PCN algorithm.]{\label{fig:sim2tree}\includegraphics[width=22em,height=17em]{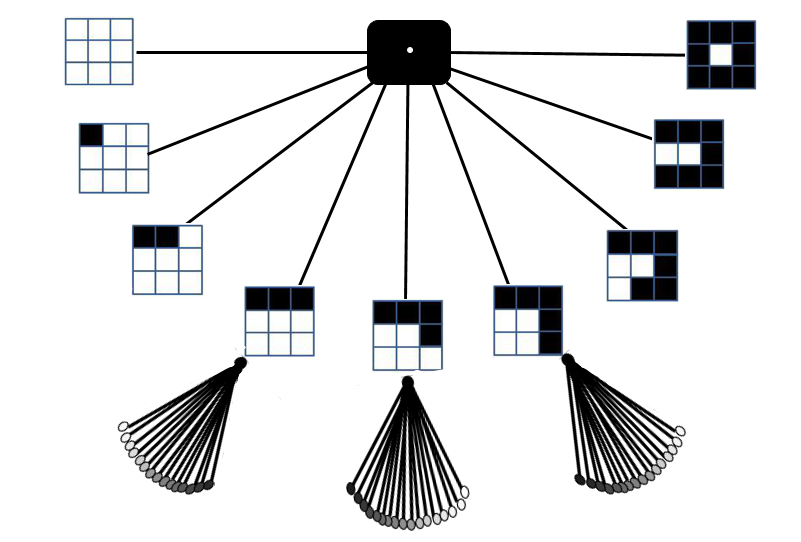} }
    \caption{Simulation results for a variable-neighborhood PCN tree with $d(\mathcal{T}_0)=2$.}%
    \label{fig:sim2}%
    \end{figure}
    
    A lattice with $150 \times 150$ sites was created after 50 iterations of the MCMC algorithm. The resulting image is presented in Figure \ref{fig:sim2sample}. The estimated tree obtained from the pruning procedure in the PCN algorithm is given in Figure \ref{fig:sim2tree}.
    
    The tree structure recovered is almost identical to the original tree $\mathcal{T}_0$ in Figure \ref{fig:sim2true}. The estimated PCN $\hat{\mathcal{T}}$ has 6 first-order contexts, like the original tree, and 48 second-order contexts, compared to the 51 contexts in $\mathcal{T}_0$. The 3 missing context neighborhoods in the second order did not appear in the generated sample. This is believed to happen due to the relatively small sample size.

    \begin{table}
    \caption{Comparison between the true probability of a site being black given the context neighborhood and the point estimate for the conditional probability in Simulation 2.}
    \label{tab:sim2}
    \begin{minipage}{0.5\textwidth}
    \centering
    \begin{tabular}{c c c }
    \toprule
    Context & True & Estimate \\
    \midrule
    \begin{minipage}{.05\textwidth}
      \includegraphics[width=2em, height=2em]{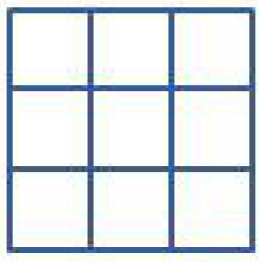}
    \end{minipage}
    & 0.3100 & 0.3844 \\[1.5mm]
    \begin{minipage}{.05\textwidth}
      \includegraphics[width=2em, height=2em]{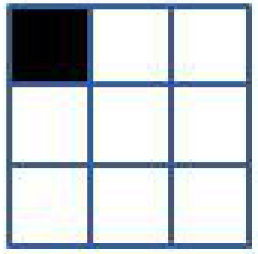}
    \end{minipage}
    & 0.3543 & 0.3567 \\[1.5mm]
    \begin{minipage}{.05\textwidth}
      \includegraphics[width=2em, height=2em]{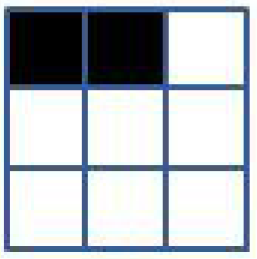}
    \end{minipage}
    & 0.4013 & 0.3736 \\[1.5mm]
    \begin{minipage}{.05\textwidth}
      \includegraphics[width=2em, height=2em]{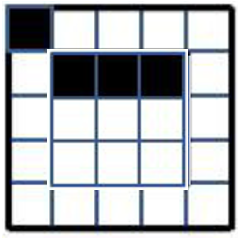}
    \end{minipage}
    & 0.1680 & 0.3000 \\[1.5mm]
    \begin{minipage}{.05\textwidth}
      \includegraphics[width=2em, height=2em]{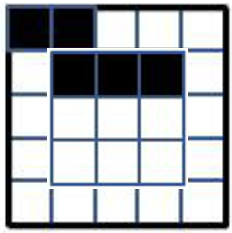}
    \end{minipage}
    & 0.1978 & 0.3103 \\[1.5mm]
    \begin{minipage}{.05\textwidth}
      \includegraphics[width=2em, height=2em]{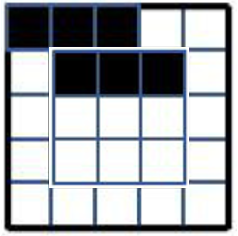}
    \end{minipage}
    & 0.2315 & 0.2965 \\[1.5mm]
    \begin{minipage}{.05\textwidth}
      \includegraphics[width=2em, height=2em]{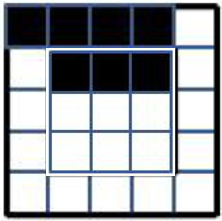}
    \end{minipage}
    & 0.2689 & 0.3205 \\[1.5mm]
    \begin{minipage}{.05\textwidth}
      \includegraphics[width=1.8em, height=1.8em]{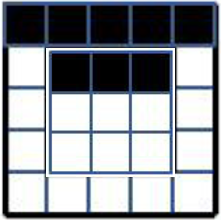}
    \end{minipage}
    & 0.3100 & 0.3462 \\[1.5mm]
    \begin{minipage}{.05\textwidth}
      \includegraphics[width=2em, height=2em]{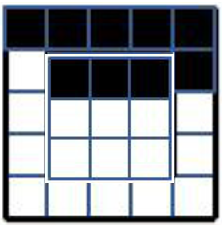}
    \end{minipage}
    & 0.3543 & 0.3448 \\[1.5mm]
    \begin{minipage}{.05\textwidth}
      \includegraphics[width=2em, height=2em]{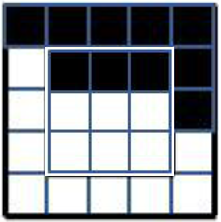}
    \end{minipage}
    & 0.4013 & 0.3889 \\[1.5mm]
    \begin{minipage}{.05\textwidth}
      \includegraphics[width=2em, height=2em]{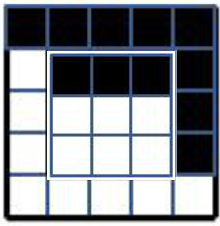}
    \end{minipage}
    & 0.4502 & 0.4601 \\[1.5mm]
    \begin{minipage}{.05\textwidth}
      \includegraphics[width=2em, height=2em]{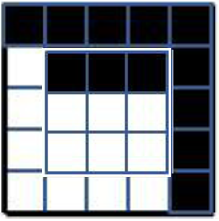}
    \end{minipage}
    & 0.5000 & 0.4944 \\[1.5mm]
    \begin{minipage}{.05\textwidth}
      \includegraphics[width=2em, height=2em]{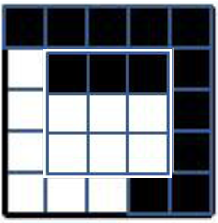}
    \end{minipage}
    & 0.5498 & 0.5372 \\[1.5mm]
    \begin{minipage}{.05\textwidth}
      \includegraphics[width=2em, height=2em]{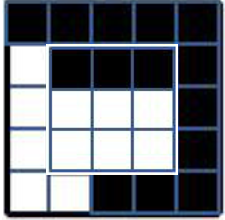}
    \end{minipage}
    & 0.5987 & 0.5633 \\[1.5mm]
    \begin{minipage}{.05\textwidth}
      \includegraphics[width=2em, height=2em]{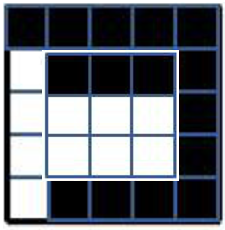}
    \end{minipage}
    & 0.6457 & 0.5876 \\[1.5mm]
    \begin{minipage}{.05\textwidth}
      \includegraphics[width=2em, height=2em]{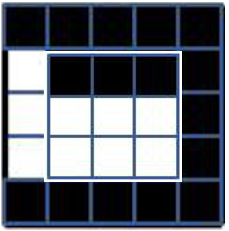}
    \end{minipage}
    & 0.6900 & 0.6316 \\[1.5mm]
    \begin{minipage}{.05\textwidth}
      \includegraphics[width=2em, height=2em]{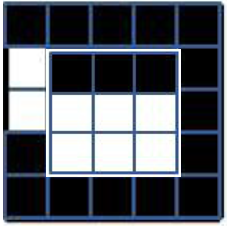}
    \end{minipage}
    & 0.7311 & 0.6111 \\[1.5mm]
    \begin{minipage}{.05\textwidth}
      \includegraphics[width=2em, height=2em]{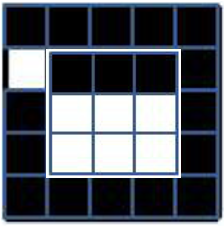}
    \end{minipage}
    & 0.7685 & 1.0000 \\[1.5mm]
    \begin{minipage}{.05\textwidth}
      \includegraphics[width=2em, height=2em]{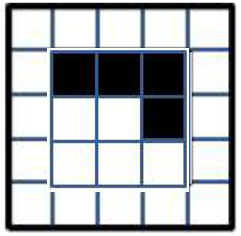}
    \end{minipage}
    & 0.168 & 0.0000 \\[1.5mm]
    \begin{minipage}{.05\textwidth}
      \includegraphics[width=2em, height=2em]{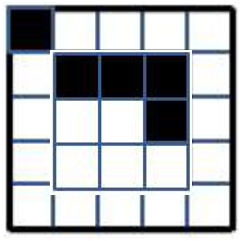}
    \end{minipage}
    & 0.1978 & 0.3333 \\[1.5mm]
    \begin{minipage}{.05\textwidth}
      \includegraphics[width=2em, height=2em]{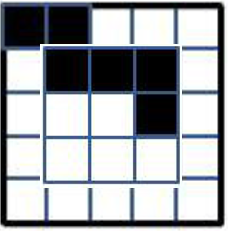}
    \end{minipage}
    & 0.2315 & 0.2973 \\[1.5mm]
    \begin{minipage}{.05\textwidth}
      \includegraphics[width=2em, height=2em]{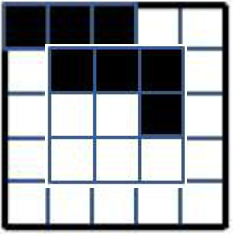}
    \end{minipage}
    & 0.2689 & 0.2673 \\[1.5mm]
    \begin{minipage}{.05\textwidth}
      \includegraphics[width=2em, height=2em]{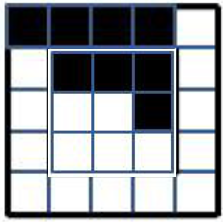}
    \end{minipage}
    & 0.3100 & 0.3515 \\[1.5mm]
    \begin{minipage}{.05\textwidth}
      \includegraphics[width=2em, height=2em]{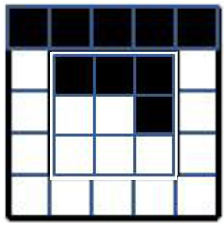}
    \end{minipage}
    & 0.3543 & 0.3830 \\[1.5mm]
    \begin{minipage}{.05\textwidth}
      \includegraphics[width=2em, height=2em]{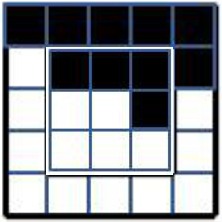}
    \end{minipage}
    & 0.4013 & 0.3970 \\[1.5mm]
    \begin{minipage}{.05\textwidth}
      \includegraphics[width=2em, height=2em]{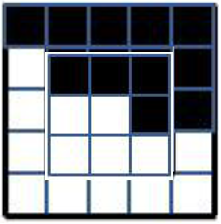}
    \end{minipage}
    & 0.4502 & 0.4648 \\[1.5mm]
    \begin{minipage}{.05\textwidth}
      \includegraphics[width=2em, height=2em]{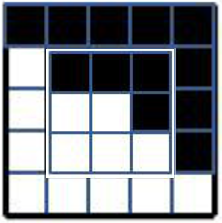}
    \end{minipage}
    & 0.5000 & 0.4908 \\
    \bottomrule
    \end{tabular}
    \end{minipage} \hfill
    \begin{minipage}{0.5\textwidth}
    \begin{tabular}{c c c }
    \toprule
    Context & True & Estimate \\
    \midrule
    \begin{minipage}{.05\textwidth}
      \includegraphics[width=2em, height=2em]{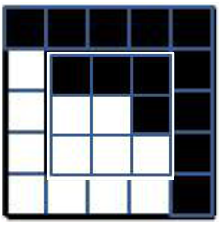}
    \end{minipage}
    & 0.5498 & 0.5342 \\[1.5mm]
    \begin{minipage}{.05\textwidth}
      \includegraphics[width=2em, height=2em]{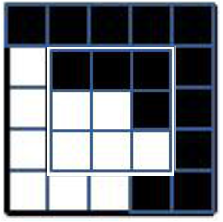}
    \end{minipage}
    & 0.5987 & 0.5623 \\[1.5mm]
    \begin{minipage}{.05\textwidth}
      \includegraphics[width=2em, height=2em]{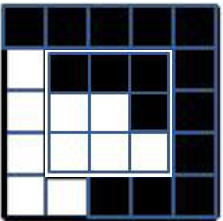}
    \end{minipage}
    & 0.6457 & 0.6034 \\[1.5mm]
    \begin{minipage}{.05\textwidth}
      \includegraphics[width=2em, height=2em]{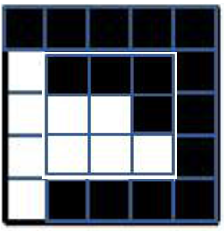}
    \end{minipage}
    & 0.6900 & 0.6302 \\[1.5mm]
    \begin{minipage}{.05\textwidth}
      \includegraphics[width=2em, height=2em]{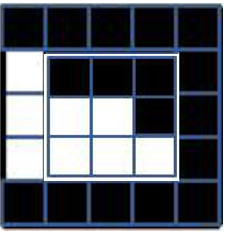}
    \end{minipage}
    & 0.7311 & 0.7216 \\[1.5mm]
    \begin{minipage}{.05\textwidth}
      \includegraphics[width=2em, height=2em]{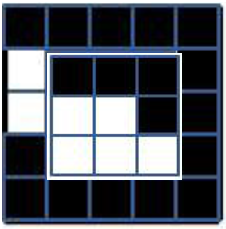}
    \end{minipage}
    & 0.7685 & 0.5946 \\[1.5mm]
    \begin{minipage}{.05\textwidth}
      \includegraphics[width=2em, height=2em]{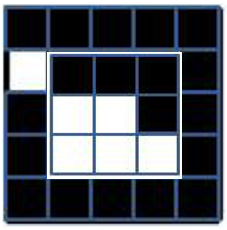}
    \end{minipage}
    & 0.8022 & 0.6000 \\[1.5mm]
    \begin{minipage}{.05\textwidth}
      \includegraphics[width=2em, height=2em]{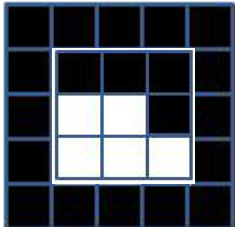}
    \end{minipage}
    & 0.8320 & 1.000 \\[1.5mm]
    \begin{minipage}{.05\textwidth}
      \includegraphics[width=2em, height=2em]{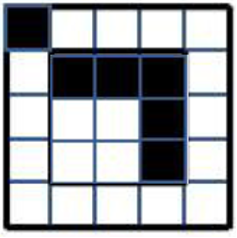}
    \end{minipage}
    & 0.2315 & 0.1667 \\[1.5mm]
    \begin{minipage}{.05\textwidth}
      \includegraphics[width=2em, height=2em]{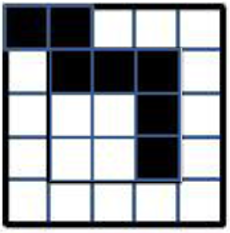}
    \end{minipage}
    & 0.2689 & 0.1176 \\[1.5mm]
    \begin{minipage}{.05\textwidth}
      \includegraphics[width=2em, height=2em]{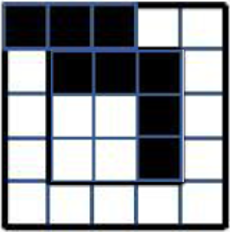}
    \end{minipage}
    & 0.3100 & 0.2424 \\[1.5mm]
    \begin{minipage}{.05\textwidth}
      \includegraphics[width=2em, height=2em]{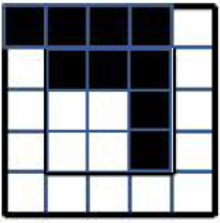}
    \end{minipage}
    & 0.3543 & 0.4043 \\[1.5mm]
    \begin{minipage}{.05\textwidth}
      \includegraphics[width=2em, height=2em]{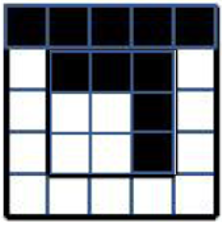}
    \end{minipage}
    & 0.4013 & 0.4646 \\[1.5mm]
    \begin{minipage}{.05\textwidth}
      \includegraphics[width=2em, height=2em]{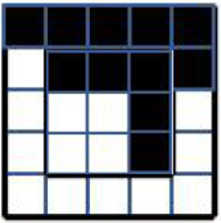}
    \end{minipage}
    & 0.4502 & 0.4987 \\[1.5mm]
    \begin{minipage}{.05\textwidth}
      \includegraphics[width=2em, height=2em]{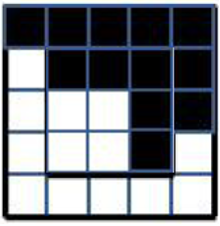}
    \end{minipage}
    & 0.5000 & 0.5139 \\[1.5mm]
    \begin{minipage}{.05\textwidth}
      \includegraphics[width=2em, height=2em]{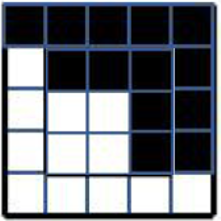}
    \end{minipage}
    & 0.5498 & 0.5822 \\[1.5mm]
    \begin{minipage}{.05\textwidth}
      \includegraphics[width=2em, height=2em]{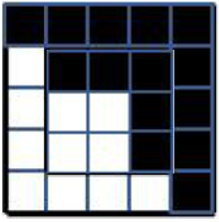}
    \end{minipage}
    & 0.5987 & 0.5997 \\[1.5mm]
    \begin{minipage}{.05\textwidth}
      \includegraphics[width=2em, height=2em]{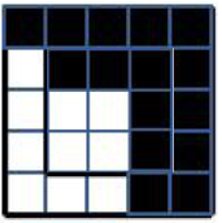}
    \end{minipage}
    & 0.6457 & 0.6460 \\[1.5mm]
    \begin{minipage}{.05\textwidth}
      \includegraphics[width=2em, height=2em]{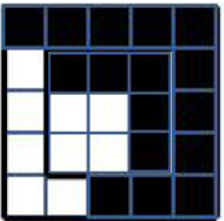}
    \end{minipage}
    & 0.6900 & 0.6774 \\[1.5mm]
    \begin{minipage}{.05\textwidth}
      \includegraphics[width=2em, height=2em]{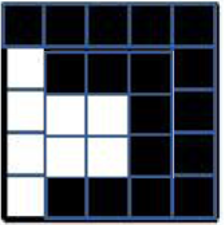}
    \end{minipage}
    & 0.7311 & 0.6889 \\[1.5mm]
    \begin{minipage}{.05\textwidth}
      \includegraphics[width=2em, height=2em]{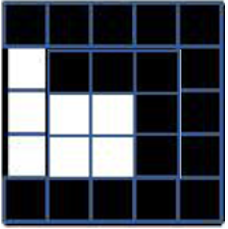}
    \end{minipage}
    & 0.7685 & 0.7153 \\[1.5mm]
    \begin{minipage}{.05\textwidth}
      \includegraphics[width=2em, height=2em]{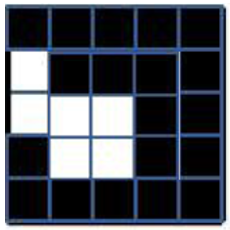}
    \end{minipage}
    & 0.8022 & 0.6061 \\[1.5mm]
    \begin{minipage}{.05\textwidth}
      \includegraphics[width=2em, height=2em]{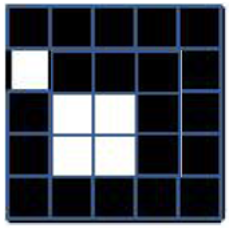}
    \end{minipage}
    & 0.8320 & 0.6154 \\[1.5mm]
    \begin{minipage}{.05\textwidth}
      \includegraphics[width=2em, height=2em]{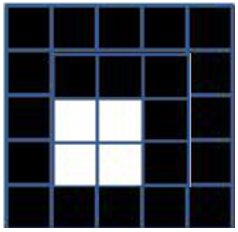}
    \end{minipage}
    & 0.8581 & 1.0000 \\[1.5mm]
    \begin{minipage}{.05\textwidth}
      \includegraphics[width=2em, height=2em]{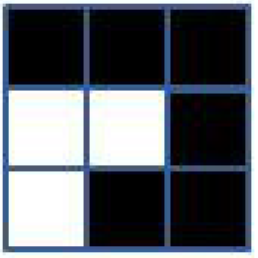}
    \end{minipage}
    & 0.5987 & 0.6118 \\[1.5mm]
    \begin{minipage}{.05\textwidth}
      \includegraphics[width=2em, height=2em]{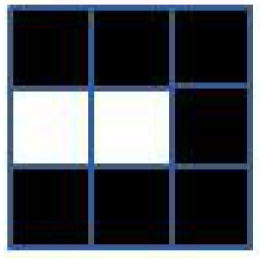}
    \end{minipage}
    & 0.6457 & 0.6387 \\[1.5mm]
    \begin{minipage}{.05\textwidth}
      \includegraphics[width=2em, height=2em]{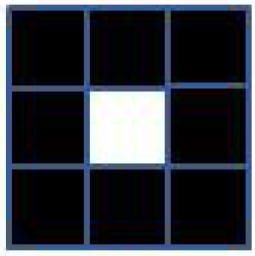}
    \end{minipage}
    & 0.6900 & 0.6952 \\
    \bottomrule
    \end{tabular}
    \end{minipage}
    \end{table}

    Table \ref{tab:sim2} shows the comparison between the conditional probabilities of the original tree and the estimated tree. 
    
    Using the same approach as in the previous simulation, we built intervals for the conditional probabilities of each context neighborhood. The 2.5\textsuperscript{th} percentile, median, and 97.5\textsuperscript{th} percentile were computed based on a sample of 81 matrices. Out of 100 matrices generated from $\mathcal{T}_0$ after 200 iterations of the MCMC algorithm, 81 of them recovered the original tree structure after the pruning procedure, and were used to build these intervals. This fluctuation is expected since there is an inherent variability within the tree structure as well as the conditional probabilities.
    
    \begin{table}
    \caption{Comparison between the true probability of a site being black given the context neighborhood and the estimated interval for each conditional probability in Simulation 2. The lower bound (LB) corresponds to the 2.5\textsuperscript{th} percentile and the upper bound (UB) is the 97.5\textsuperscript{th}  percentile.}
    \label{tab:sim2int}
    \makebox[1.1\linewidth]{
    \begin{minipage}{0.55\linewidth}
    \begin{tabular}{c c c c c }
    \toprule
    \multirow{2}{*}{Context} &
    \multirow{2}{*}{True} &
    \multicolumn{3}{c}{Interval Estimate } \\
    \cline{3-5}
    & & LB & Median & UB \\
    \midrule
    \begin{minipage}{.05\textwidth}
      \includegraphics[width=1.8em, height=1.8em]{Figures/0pretos.PNG}
    \end{minipage}
    & 0.3100 & 0.2524 & 0.3265 & 0.3968 \\[1mm]
    \begin{minipage}{.05\textwidth}
      \includegraphics[width=1.8em, height=1.8em]{Figures/1preto.PNG}
    \end{minipage}
    & 0.3543 & 0.3305 & 0.3575 & 0.3822 \\[1mm]
    \begin{minipage}{.05\textwidth}
      \includegraphics[width=1.8em, height=1.8em]{Figures/2pretos.PNG}
    \end{minipage}
    & 0.4013 & 0.3634 & 0.3797 & 0.4075 \\[1mm]
    \begin{minipage}{.05\textwidth}
      \includegraphics[width=1.8em, height=1.8em]{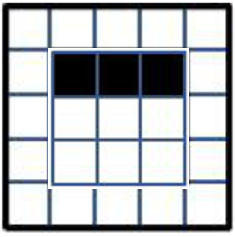}
    \end{minipage}
    & 0.1419 & 0.0000 & 0.0000 & 1.0000 \\[1mm]
    \begin{minipage}{.05\textwidth}
      \includegraphics[width=1.8em, height=1.8em]{Figures/1e3pretos.PNG}
    \end{minipage}
    & 0.1680 & 0.0000 & 0.3000 & 0.5682 \\[1mm]
    \begin{minipage}{.05\textwidth}
      \includegraphics[width=1.8em, height=1.8em]{Figures/2e3pretos.PNG}
    \end{minipage}
    & 0.1978 & 0.1556 & 0.2826 & 0.4285 \\[1mm]
    \begin{minipage}{.05\textwidth}
      \includegraphics[width=1.8em, height=1.8em]{Figures/3e3pretos.PNG}
    \end{minipage}
    & 0.2315 & 0.2416 & 0.3089 & 0.3982 \\[1mm]
    \begin{minipage}{.05\textwidth}
      \includegraphics[width=1.8em, height=1.8em]{Figures/4e3pretos.PNG}
    \end{minipage}
    & 0.2689 & 0.2704 & 0.3297 & 0.3856 \\[1mm]
    \begin{minipage}{.05\textwidth}
      \includegraphics[width=1.8em, height=1.8em]{Figures/5e3pretos.PNG}
    \end{minipage}
    & 0.3100 & 0.3030 & 0.3496 & 0.4015 \\[1mm]
    \begin{minipage}{.05\textwidth}
      \includegraphics[width=1.8em, height=1.8em]{Figures/6e3pretos.PNG}
    \end{minipage}
    & 0.3543 & 0.3409 & 0.3808 & 0.4103 \\[1mm]
    \begin{minipage}{.05\textwidth}
      \includegraphics[width=1.8em, height=1.8em]{Figures/7e3pretos.PNG}
    \end{minipage}
    & 0.4013 & 0.3736 & 0.4103 & 0.4512 \\[1mm]
    \begin{minipage}{.05\textwidth}
      \includegraphics[width=1.8em, height=1.8em]{Figures/8e3pretos.PNG}
    \end{minipage}
    & 0.4502 & 0.4054 & 0.4446 & 0.4845 \\[1mm]
    \begin{minipage}{.05\textwidth}
      \includegraphics[width=1.8em, height=1.8em]{Figures/9e3pretos.PNG}
    \end{minipage}
    & 0.5000 & 0.4327 & 0.4778 & 0.5202 \\[1mm]
    \begin{minipage}{.05\textwidth}
      \includegraphics[width=1.8em, height=1.8em]{Figures/10e3pretos.PNG}
    \end{minipage}
    & 0.5498 & 0.4690 & 0.5185 & 0.5858 \\[1mm]
    \begin{minipage}{.05\textwidth}
      \includegraphics[width=1.8em, height=1.8em]{Figures/11e3pretos.PNG}
    \end{minipage}
    & 0.5987 & 0.4727 & 0.5566 & 0.6253 \\[1mm]
    \begin{minipage}{.05\textwidth}
      \includegraphics[width=1.8em, height=1.8em]{Figures/12e3pretos.PNG}
    \end{minipage}
    & 0.6457 & 0.4737 & 0.5966 & 0.6984 \\[1mm]
    \begin{minipage}{.05\textwidth}
      \includegraphics[width=1.8em, height=1.8em]{Figures/13e3pretos.PNG}
    \end{minipage}
    & 0.6900 & 0.5076 & 0.6429 & 0.8028 \\[1mm]
    \begin{minipage}{.05\textwidth}
      \includegraphics[width=1.8em, height=1.8em]{Figures/14e3pretos.PNG}
    \end{minipage}
    & 0.7311 & 0.4210 & 0.6667 & 0.9756 \\[1mm]
    \begin{minipage}{.05\textwidth}
      \includegraphics[width=1.8em, height=1.8em]{Figures/15e3pretos.PNG}
    \end{minipage}
    & 0.7685 & 0.0000 & 0.7500 & 1.0000 \\[1mm]
    \begin{minipage}{.05\textwidth}
      \includegraphics[width=1.8em, height=1.8em]{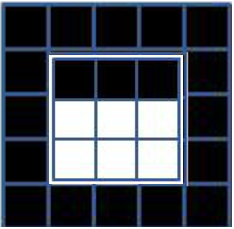}
    \end{minipage}
    & 0.8022 & 0.0000 & 1.0000 & 1.0000 \\[1mm]
    \begin{minipage}{.05\textwidth}
      \includegraphics[width=1.8em, height=1.8em]{Figures/0e4pretos.PNG}
    \end{minipage}
    & 0.168 & 0.0000 & 0.0000 & 1.0000 \\[1mm]
    \begin{minipage}{.05\textwidth}
      \includegraphics[width=1.8em, height=1.8em]{Figures/1e4pretos.PNG}
    \end{minipage}
    & 0.1978 &  0.0000 & 0.2000 & 0.5656 \\[1mm]
    \begin{minipage}{.05\textwidth}
      \includegraphics[width=1.8em, height=1.8em]{Figures/2e4pretos.PNG}
    \end{minipage}
    & 0.2315 & 0.1157 & 0.2857 & 0.4818 \\[1mm]
    \begin{minipage}{.05\textwidth}
      \includegraphics[width=1.8em, height=1.8em]{Figures/3e4pretos.PNG}
    \end{minipage}
    & 0.2689 & 0.2104 & 0.3220 & 0.4435 \\[1mm]
    \begin{minipage}{.05\textwidth}
      \includegraphics[width=1.8em, height=1.8em]{Figures/4e4pretos.PNG}
    \end{minipage}
    & 0.3100 & 0.2987 & 0.3557 & 0.4310 \\[1mm]
    \begin{minipage}{.05\textwidth}
      \includegraphics[width=1.8em, height=1.8em]{Figures/5e4pretos.PNG}
    \end{minipage}
    & 0.3543 & 0.3220 & 0.3922 & 0.4363 \\[1mm]
    \begin{minipage}{.05\textwidth}
      \includegraphics[width=1.8em, height=1.8em]{Figures/6e4pretos.PNG}
    \end{minipage}
    & 0.4013 & 0.3975 & 0.4231 & 0.4748 \\[1mm]
    \begin{minipage}{.05\textwidth}
      \includegraphics[width=1.8em, height=1.8em]{Figures/7e4pretos.PNG}
    \end{minipage}
    & 0.4502 & 0.4206 & 0.4673 & 0.5071 \\[1mm]
    \begin{minipage}{.05\textwidth}
      \includegraphics[width=1.8em, height=1.8em]{Figures/8e4pretos.PNG}
    \end{minipage}
    & 0.5000 & 0.4648 & 0.5032 & 0.5389 \\
    \bottomrule
    \end{tabular}
    \end{minipage}\hfill
    \begin{minipage}{0.55\linewidth}
    \vspace{-2em}
    \begin{tabular}{c c c c c}
    \toprule
    \multirow{2}{*}{Context} &
    \multirow{2}{*}{True} &
    \multicolumn{3}{c}{Interval Estimate } \\
    \cline{3-5}
    & & LB & Median & UB \\
    \midrule
    \begin{minipage}{.05\textwidth}
      \includegraphics[width=1.8em, height=1.8em]{Figures/9e4pretos.PNG}
    \end{minipage}
    & 0.5498 & 0.5095 & 0.5439 & 0.5730 \\[1mm]
    \begin{minipage}{.05\textwidth}
      \includegraphics[width=1.8em, height=1.8em]{Figures/10e4pretos.PNG}
    \end{minipage}
    & 0.5987 & 0.5351 & 0.5764 & 0.6207 \\[1mm]
    \begin{minipage}{.05\textwidth}
      \includegraphics[width=1.8em, height=1.8em]{Figures/11e4pretos.PNG}
    \end{minipage}
    & 0.6457 & 0.5678 & 0.6063 & 0.6468 \\[1mm]
    \begin{minipage}{.05\textwidth}
      \includegraphics[width=1.8em, height=1.8em]{Figures/12e4pretos.PNG}
    \end{minipage}
    & 0.6900 & 0.5778 & 0.6407 & 0.7078 \\[1mm]
    \begin{minipage}{.05\textwidth}
      \includegraphics[width=1.8em, height=1.8em]{Figures/13e4pretos.PNG}
    \end{minipage}
    & 0.7311 & 0.5588 & 0.6667 & 0.7943 \\[1mm]
    \begin{minipage}{.05\textwidth}
      \includegraphics[width=1.8em, height=1.8em]{Figures/14e4pretos.PNG}
    \end{minipage}
    & 0.7685 & 0.5059 & 0.6757 & 0.8505 \\[1mm]
    \begin{minipage}{.05\textwidth}
      \includegraphics[width=1.8em, height=1.8em]{Figures/15e4pretos.PNG}
    \end{minipage}
    & 0.8022 & 0.3486 & 0.6667 & 1.0000 \\[1mm]
    \begin{minipage}{.05\textwidth}
      \includegraphics[width=1.8em, height=1.8em]{Figures/16e4pretos.PNG}
    \end{minipage}
    & 0.8320 & 0.0000 & 1.0000 & 1.0000 \\[1mm]
    \begin{minipage}{.05\textwidth}
      \includegraphics[width=1.8em, height=1.8em]{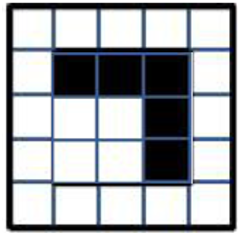}
    \end{minipage}
    & 0.1978 & 0.0000 & 0.0000 & 1.0000 \\[1mm]
    \begin{minipage}{.05\textwidth}
      \includegraphics[width=1.8em, height=1.8em]{Figures/1e5pretos.PNG}
    \end{minipage}
    & 0.2315 & 0.0000 & 0.2679 & 1.0000 \\[1mm]
    \begin{minipage}{.05\textwidth}
      \includegraphics[width=1.8em, height=1.8em]{Figures/2e5pretos.PNG}
    \end{minipage}
    & 0.2689 & 0.0935 & 0.3333 & 0.5837 \\[1mm]
    \begin{minipage}{.05\textwidth}
      \includegraphics[width=1.8em, height=1.8em]{Figures/3e5pretos.PNG}
    \end{minipage}
    & 0.3100 & 0.2140 & 0.3542 & 0.5371 \\[1mm]
    \begin{minipage}{.05\textwidth}
      \includegraphics[width=1.8em, height=1.8em]{Figures/4e5pretos.PNG}
    \end{minipage}
    & 0.3543 & 0.3260 & 0.4086 & 0.4796 \\[1mm]
    \begin{minipage}{.05\textwidth}
      \includegraphics[width=1.8em, height=1.8em]{Figures/5e5pretos.PNG}
    \end{minipage}
    & 0.4013 & 0.3615 & 0.4417 & 0.5020 \\[1mm]
    \begin{minipage}{.05\textwidth}
      \includegraphics[width=1.8em, height=1.8em]{Figures/6e5pretos.PNG}
    \end{minipage}
    & 0.4502 & 0.4200 & 0.4836 & 0.5399 \\[1mm]
    \begin{minipage}{.05\textwidth}
      \includegraphics[width=1.8em, height=1.8em]{Figures/7e5pretos.PNG}
    \end{minipage}
    & 0.5000 & 0.4795 & 0.5188 & 0.5613 \\[1mm]
    \begin{minipage}{.05\textwidth}
      \includegraphics[width=1.8em, height=1.8em]{Figures/8e5pretos.PNG}
    \end{minipage}
    & 0.5498 & 0.5153 & 0.5605 & 0.5908 \\[1mm]
    \begin{minipage}{.05\textwidth}
      \includegraphics[width=1.8em, height=1.8em]{Figures/9e5pretos.PNG}
    \end{minipage}
    & 0.5987 & 0.5576 & 0.5906 & 0.6179 \\[1mm]
    \begin{minipage}{.05\textwidth}
      \includegraphics[width=1.8em, height=1.8em]{Figures/10e5pretos.PNG}
    \end{minipage}
    & 0.6457 & 0.5872 & 0.6230 & 0.6631 \\[1mm]
    \begin{minipage}{.05\textwidth}
      \includegraphics[width=1.8em, height=1.8em]{Figures/11e5pretos.PNG}
    \end{minipage}
    & 0.6900 & 0.6063 & 0.6481 & 0.6977 \\[1mm]
    \begin{minipage}{.05\textwidth}
      \includegraphics[width=1.8em, height=1.8em]{Figures/12e5pretos.PNG}
    \end{minipage}
    & 0.7311 & 0.6129 & 0.6759 & 0.7218 \\[1mm]
    \begin{minipage}{.05\textwidth}
      \includegraphics[width=1.8em, height=1.8em]{Figures/13e5pretos.PNG}
    \end{minipage}
    & 0.7685 & 0.5985 & 0.7086 & 0.7663 \\[1mm]
    \begin{minipage}{.05\textwidth}
      \includegraphics[width=1.8em, height=1.8em]{Figures/14e5pretos.PNG}
    \end{minipage}
    & 0.8022 & 0.5682 & 0.7234 & 0.8694 \\[1mm]
    \begin{minipage}{.05\textwidth}
      \includegraphics[width=1.8em, height=1.8em]{Figures/15e5pretos.PNG}
    \end{minipage}
    & 0.8320 & 0.5000 & 0.7143 & 0.9771 \\[1mm]
    \begin{minipage}{.05\textwidth}
      \includegraphics[width=1.8em, height=1.8em]{Figures/16e5pretos.PNG}
    \end{minipage}
    & 0.8581 & 0.0000 & 0.7083 & 1.0000 \\[1mm]
    \begin{minipage}{.05\textwidth}
      \includegraphics[width=1.8em, height=1.8em]{Figures/6pretos.PNG}
    \end{minipage}
    & 0.5987 & 0.5973 & 0.6162 & 0.6414 \\[1mm]
    \begin{minipage}{.05\textwidth}
      \includegraphics[width=1.8em, height=1.8em]{Figures/7pretos.PNG}
    \end{minipage}
    & 0.6457 & 0.6194 & 0.6468 & 0.6723 \\[1mm]
    \begin{minipage}{.05\textwidth}
      \includegraphics[width=1.8em, height=1.8em]{Figures/8pretos.PNG}
    \end{minipage}
    & 0.6900 & 0.6030 & 0.6839 & 0.7427 \\
    \bottomrule
    \end{tabular}
    \end{minipage}}
    \end{table}

    The results of the interval estimation for the conditional probabilities of Simulation 2 are presented in Table \ref{tab:sim2int}. 
    
    The estimated intervals show a reasonable empirical coverage of the true values of the probabilities, around $93\%$. The range of an interval varied depending on the number of times a context neighborhood was observed within the samples analyzed and how many samples had that specific configuration. Due to low frequencies for eight context neighborhoods (appearing less than ten times within a matrix), the resulting interval covered the entire parametric space.

\subsubsection{Simulation 2: Second-order PCN {$\mathcal{T}_0$}}
\label{subsec:sim3}

    The second simulation was created to analyze the performance of the PCN algorithm applied to a sample of a complete second-order PCN tree $\mathcal{T}_0$. As given by Equation (\ref{eq:leaves}), the full second-order tree has 153 context neighborhoods. That is, each first-order node stemming from the root has 17 children nodes, and all of them are considered context neighborhoods.
    
    \begin{figure}[h]%
    \centering
    \subfloat[
    Sample lattice from a second-order PCN $\mathcal{T}_0$.]{\label{fig:sim3sample}\includegraphics[width=16em, height=16em]{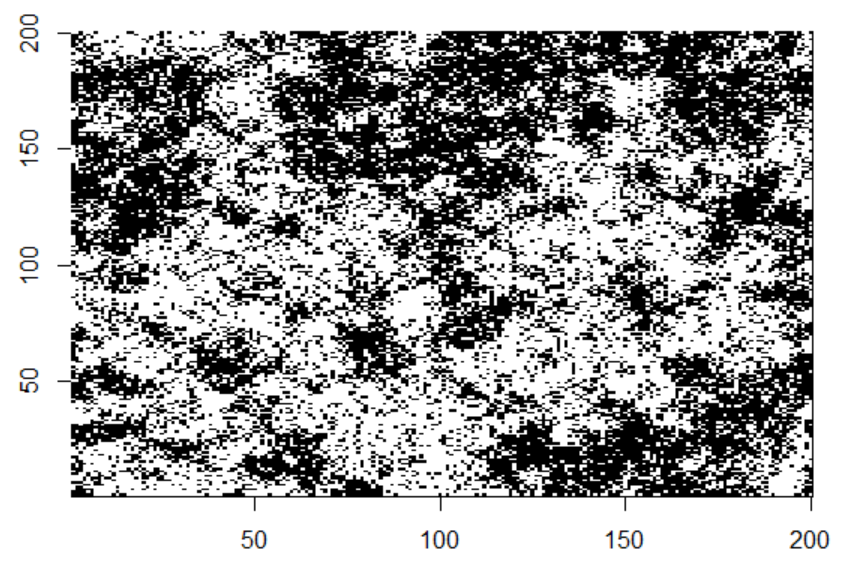} }%
    \qquad
    \subfloat[
    Estimated PCN $\hat{\mathcal{T}}$ after running the PCN algorithm.]{\label{fig:sim3tree}\includegraphics[width=22em,height=16em]{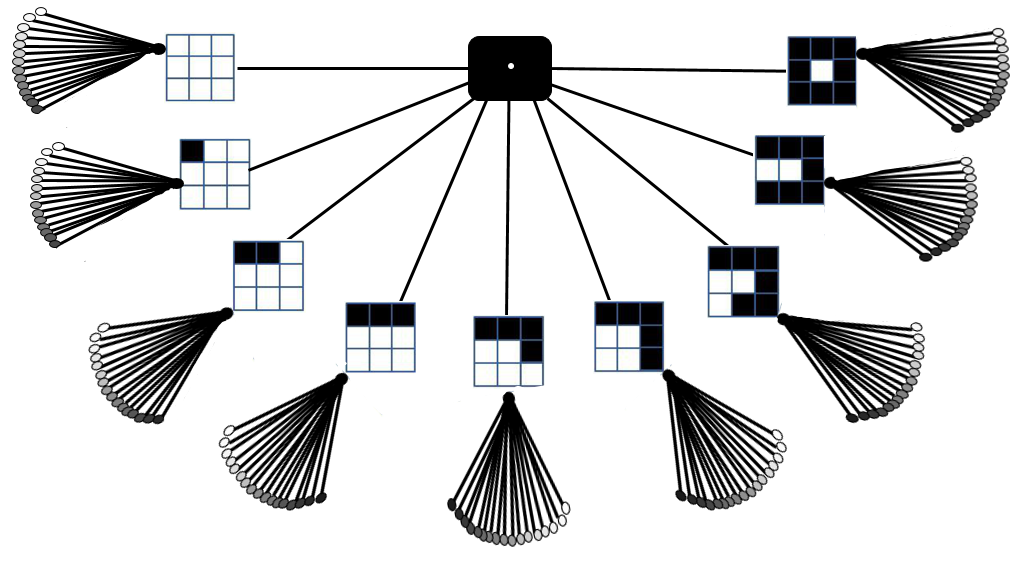} }
    \caption{Simulation results for a complete second-order PCN tree.}%
    \label{fig:sim3}%
    \end{figure}
    
    Figure \ref{fig:sim3sample} presents the $200\times200$ matrix simulated after 100 iterations of the sampling algorithm. Figure \ref{fig:sim3tree} shows the estimated PCN tree obtained through the PCN algorithm.
    
    As before, the structure of the estimated tree $\hat{\mathcal{T}}$ is quite similar to the true tree $\mathcal{T}_0$. However, the estimated tree contains a total of 141 context neighborhoods of second order rather than 153. Like in Simulation 1, the 12 missing contexts did not appear in the sample under study and, therefore, did not show up in $\hat{\mathcal{T}}$. A larger lattice would be necessary to capture all possible second-order frame configurations.
    
    Due to the large number of leaves within this tree, we chose to omit the comparisons between the true conditional probabilities of each context neighborhood and their estimated values. 
    
    We built an interval for the estimated conditional probabilities of this process, based on a sample of 50 matrices. The matrices were generated after 400 iterations of the MCMC algorithm and selected after correctly recovering the PCN tree structure. We created 149 interval estimates for the 153 total conditional probabilities of PCN $\mathcal{T}_0$. Instead of intervals, we provided point estimates for two context neighborhoods since those configurations were each observed once inside one matrix. 2 neighborhoods did not appear in a single matrix. Hence, no estimate was provided. All estimated intervals contained the true conditional probability. In 20 of them, however, the range covered the entire parametric space due to the extremely low counts for those particular context neighborhoods. 
    
    The scenarios presented in this section were run using three distinct machines. Generating a single matrix in Simulation 2 
    took approximately 16 hours.
    Subsequently, the PCN algorithm was run for approximately 25 minutes for the matrix in Simulation 2.
    
    
    These times were recorded for a computer with an Intel i5 processor running at 1.6 GHz and using 4GB of RAM. Creating a sample of matrices in Simulation 2 and applying the PCN algorithm to each matrix took approximately 33 hours. This task was performed with a more powerful machine available at {UFMG}'s Spatial Statistics Laboratory, which has an Intel Xeon processor running at 3.7GHz and using 128GB of RAM. 

We observe that the results in \cite{CSISZAR2006b} were proved for a neighborhood with sides given by $D(n) = (\log |\Lambda_n|)^{\frac{1}{4}}$. However, their findings did not address the issue of tree estimation. Based on our simulation results, we believe that this limit could be enhanced.

   All scripts used are available at:  https://github.com/denisedsma/PCN.

    \section{Recovering Spatial Dependency of Fires in the Pantanal Biome}
    \label{sec:app}
    
    The previous section showed the adequacy of the PCN model and algorithm through simulation studies. Now we seek to demonstrate an application of this methodology to a real-world dataset. 
    
    Motivated by the record number of fire foci in the Pantanal Biome in the Center-West Region of Brazil through September 2020 \cite{INPEfocos}, we conducted a study on the spatial dependency of fires in that region. Fires cause damage to local biodiversity, increase CO\textsubscript{2} emissions, and can severely affect people's health. The PCN model can provide insight into the spatial dependency structure of this phenomenon, as well as quantify the conditional probabilities of this unknown process. This type of information can be valuable to shape a more efficient fire prevention plan. 
    
\subsection{MODIS Data}
\label{sec:modis}
    
    We chose to work with NASA's Moderate Resolution Imaging Spectroradiometer ({MODIS}) Burned Area product due to its reliability and the fact that it is a well-documented data source. The MCD64A1 Burned Area Product is a monthly and gridded 500-meter product containing burned areas per pixel. Therefore, we can evaluate the pixels in the grid as we evaluated the sites of a lattice in our simulation study in Section \ref{sec:sim}.
  
   
    All the results presented in this section were obtained through the MCD64A1 GeoTIFF files. These files are divided into 24 different windows. We selected burned area product data for Windows 5 and 6 in September 2020. This was done by downloading the GeoTIFF files from the fuoco {SFTP} server as directed by the MODIS Burned Area Product User's Guide \cite{MODIS}.
  
    We will disregard the temporal component of this study and focus only on its spatial aspect. The PCN model will be seen as a representation of a Markovian process for a given moment. We are interested in investigating the spatial dependency of fires in Pantanal in an unprecedented time in history. September of 2020 saw 8106 fires detected by the reference satellite, compared to 2887 for the same month in the prior year. Before that, the maximum number of fire foci was 5993 recorded in August of 2005 \cite{INPEfocos}.

 The data matrix contains three categories: fire, unburned land, and water. Minor modifications were made for compatibility with the PCN algorithm.

 We observe that the scripts used to obtain the data matrix are also available at https://github.com/denisedsma/PCN. 
    
    \begin{figure}
        \centering
        \includegraphics[width=25em,height=25em]{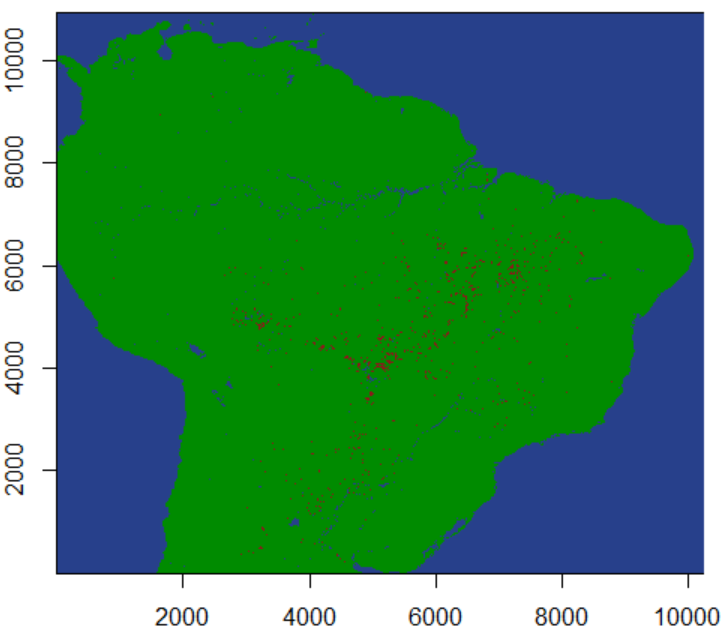}
    \caption{Matrix object corresponding to the MCD64A1 Burned Area Product for Windows 5 and 6 regarding September of 2020. A green pixel represents unburned land, a blue pixel corresponds to water, and a red pixel represents fire.}
    \label{fig:amostraBR}
    \end{figure}
    
    Figure \ref{fig:amostraBR} displays the MCD64A1 Burned Area Product for September 2020 corresponding to Windows 5 and 6 after the above steps were performed. Each category is illustrated by a different color pixel. Fires are red, unburned land is green, and water is blue.
    
    \begin{figure}
        \centering
        \includegraphics[width=30em,height=25em]{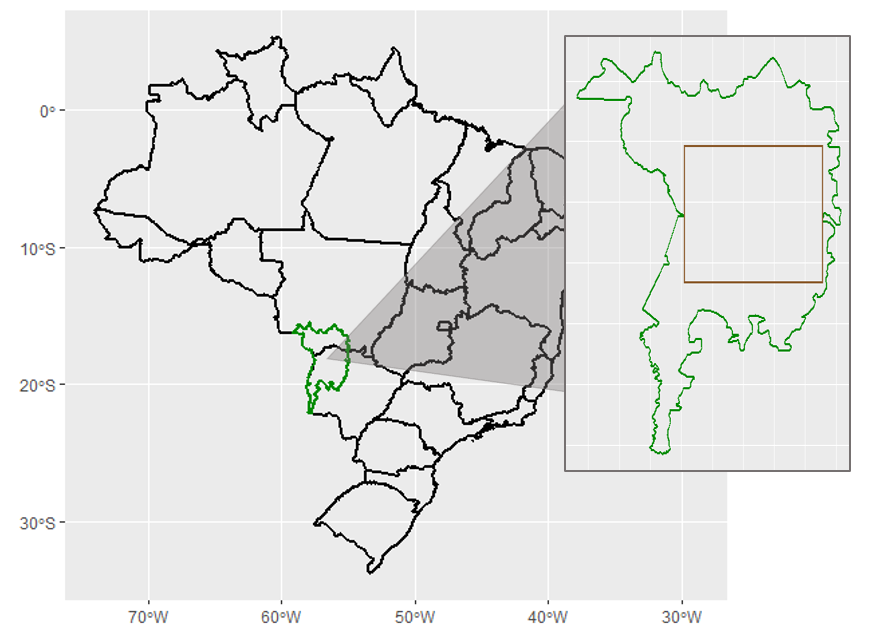}
    \caption{Map of Brazil divided by its states, created from a shapefile obtained from \cite{IBGE}. The Pantanal biome boundary is represented in green. The brown box inside the Pantanal region corresponds to the sample selected for the analysis. }
    \label{fig:box}
    \end{figure}
    
    Next, using the \texttt{rgdal} R package \cite{RGDAL} and a shapefile obtained from \cite{INPEshp}, we examined the boundaries of the Pantanal biome. Based on these geographic coordinates, we selected the largest square matrix within Pantanal to analyze. Figure \ref{fig:box} shows the location of Pantanal (in green) inside the map of Brazil. The brown square inside the Pantanal boundary represents the sample under study.
    
    \begin{figure}
        \centering
        \includegraphics[width=24em,height=24em]{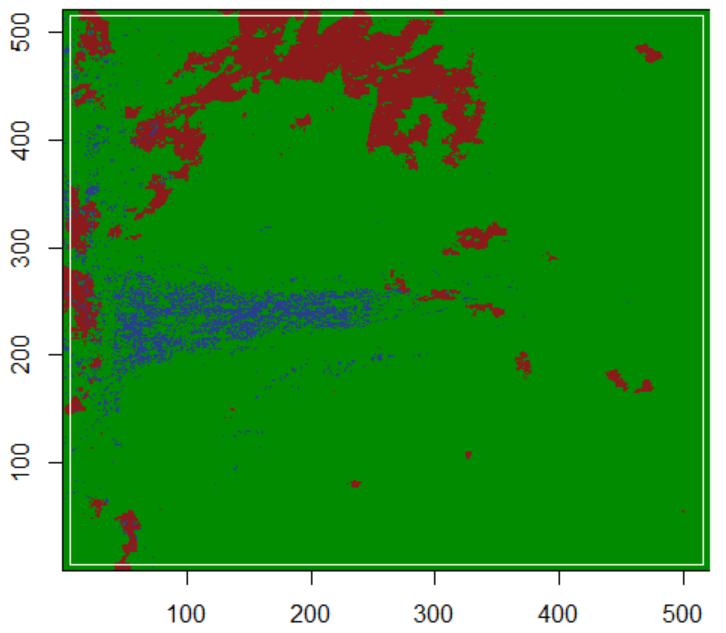}
    \caption{Sample matrix of the Pantanal region, including the sites outside the border considered in the PCN algorithm. The color scheme is the same as before: unburned land is green, water is blue, and fire is red. The region inside the white box is the $510\times510$ matrix evaluated by the PCN model.}
    \label{fig:amostraPant}
    \end{figure}
    
    The final sample is a $510\times510$ matrix as displayed in Figure \ref{fig:amostraPant}. There are a total of 260,100 pixels of which 230,114 are unburned land, 7,881 are water, and 22,105 are fire. Although there are three possible values for a site, when running the PCN algorithm, we consider a binary alphabet in our formulas. This is due to the fact that we are studying the dependency structure of fires. Water pixels will remain water pixels regardless of their neighborhood, therefore, it does not make sense to study the conditional probability of those sites becoming fire. So, for the purpose of the PCN model, there are only two possibilities for a site: fire and not fire. Since water sites are not dependent on the context neighborhood, they are not evaluated or counted in the PCN algorithm. They only influence this process when present in the neighborhood of a ``valid" site. Then, water pixels are counted as ``not fire" along with unburned land pixels.
    
\subsection{Results}
\label{sec:pantanalRes}

    The PCN algorithm used in the simulation study had to be modified to produce results for real-world data analysis. For the reason specified earlier, we had to make adjustments to skip the neighborhood evaluation of water pixels inside the sample. This way, water sites, and their neighborhood configurations were not counted as part of this unknown process. We used the real values outside the selected sample as a buffer.
    
    The most time-consuming stage of the algorithm builds a tree from the sample under study containing all the site counts as well as their neighborhood counts. In the simulation study, building this tree for a $200\times200$ matrix took approximately 32 minutes. In the real-world application study, the same step was performed in 4 minutes for a $510\times510$ matrix, despite the depth of the tree growing with the sample size. It is worth noting that the other stages of the PCN algorithm, responsible for calculating $\tilde P_{\mathcal{D}^{j}} \left(a(\Lambda_n)\right)$, the value $V_{\mathcal D^{j}}^{D}(a(\Lambda_n))$ and the indicator $\chi_{\mathcal D^{j}}^D(a(\Lambda_n))$, as well as pruning the tree, only took a few seconds to run in both studies. The recorded times were observed on a computer with an Intel i7 processor running at 1.3 GHz and using 12GB of RAM. 

\subsubsection{PCN {$\hat{\mathcal{T}}$}}
\label{subsec:pantanalT}

    \begin{figure}
    \centering
        \includegraphics[width=42em,height=17em]{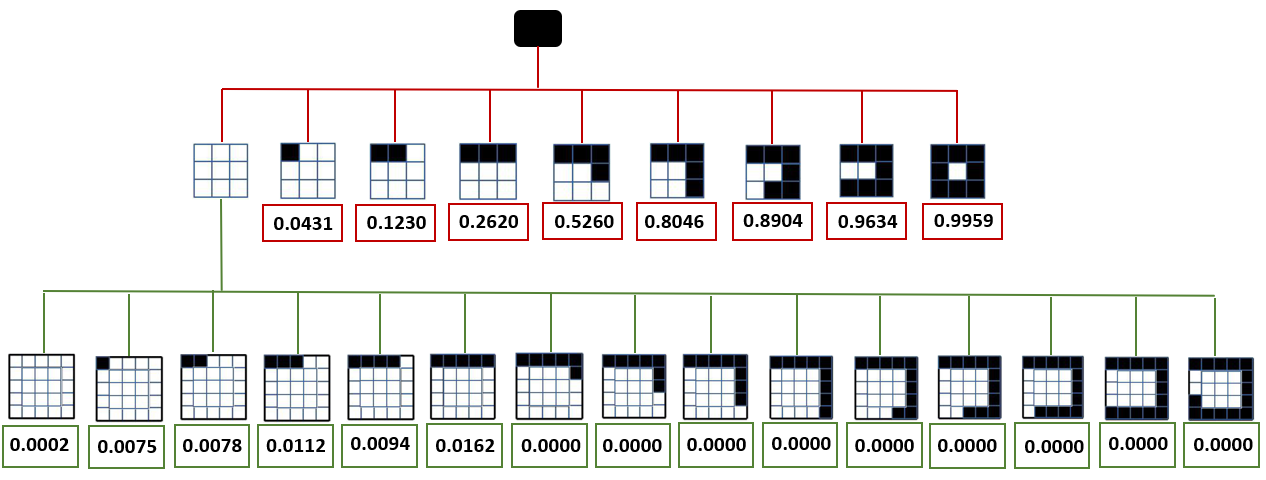}
    \caption{PCN $\hat{\mathcal{T}}$ recovered from the PCN algorithm applied to the Pantanal matrix. The point estimate for the conditional probability of each context (or leaf) is given underneath the neighborhood configuration. Red boxes refer to first-order contexts whereas green boxes refer to second-order contexts. This tree represents the probability of a site being fire given the neighborhood.}
    \label{fig:treePant}
    \end{figure}

    The resulting PCN tree and the estimated conditional probabilities of this process are given by Figure \ref{fig:treePant}. Sites inside the sample are either fire (black) or ``not fire" (white). As demonstrated by the root, this PCN tree represents the spatial dependency structure and probabilities of a site being fire conditioned on the context neighborhood.
    
    Figure \ref{fig:treePant} indicates that there are 23 total context neighborhoods. Every first-order neighborhood configuration is a context for this process, except for the neighborhood with 8 white sites in the first frame. In other words, if no fires were observed in the first-order neighborhood, we need to inspect the second-order neighborhood to determine the conditional probability of the site under study. In addition, there are 15 second-order context neighborhoods out of 17 possible second-order configurations. First-order frames with 0 black sites combined with second-order frames with 15 and 16 black sites did not occur in the sample analyzed and, therefore, did not appear in the estimated tree. Also, contexts with 8 to 14 black sites in the second frame appeared less than 30 times in the sample and resulted in an estimated conditional probability equal to zero.
    
    In general, having sites of fire in the neighborhood increases the probability of the center site being fire. Also, the conditional probability of fires in Pantanal is mostly dependent on the immediate neighbors experiencing fires. In the cases where that does not happen, the conditional probabilities are determined based on a larger neighborhood scope, the second-order neighborhood.

\subsubsection{Building Interval Estimates via Bootstrap}
\label{subsec:bootstrap}
 We applied the bootstrap method described in Section \ref{bootIC} to build confidence intervals for the conditional probabilities resampling from the estimated PCN $\hat{\mathcal{T}}$ given in Figure \ref{fig:treePant}.
    
    \begin{figure}
    \centering
        \includegraphics[width=42em,height=25em]{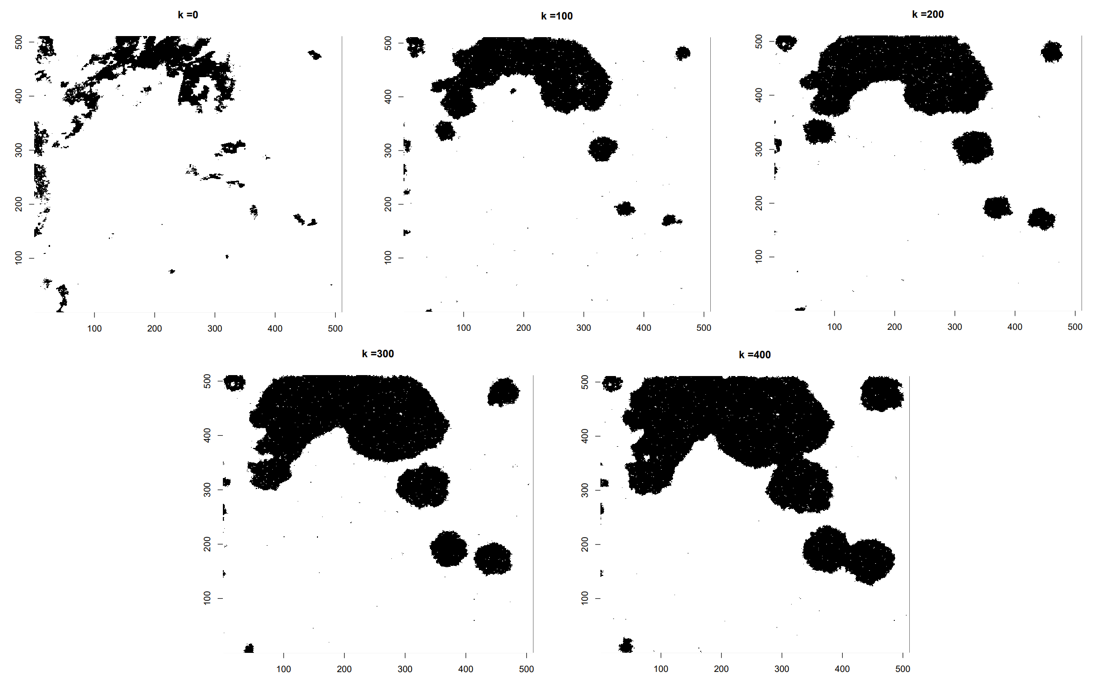}
    \caption{Iterations $k=0, 100, 200, 300,400$ of the MCMC algorithm.}
    \label{fig:k}
    \end{figure}
    
    A conditional probability of being black is attributed to a site based on its neighbors, as dictated by the PCN tree $\mathcal{T}_0$. An acceptance step, similar to the Metropolis-Hastings algorithm \cite{METROPOLIS1953,HASTINGS1970}, is then used to determine whether the site under evaluation is black or white. Once this procedure is done for all sites, we have completed the first iteration. We perform iterations until the image ``stabilizes". From then on, we consider that the process has converged to the target distribution. adjustments.

    We used the real matrix displayed in Figure \ref{fig:amostraPant} as the starting point. Water sites did not suffer any changes throughout the iterations since they do not belong to the process we are trying to estimate. Also, the sampling algorithm could not run without a value for the conditional probabilities of the 2 ``missing" second-order contexts. So, in the acceptance step, we used the empirical probability of a site being black-conditioned on 8 white sites in the first frame. 
    
    \begin{figure}%
    \centering
    \subfloat[
    Difference between the proportion of frames with 4 white sites in the first frame across each iteration.]{\label{fig:diff4}\includegraphics[scale = 0.42]{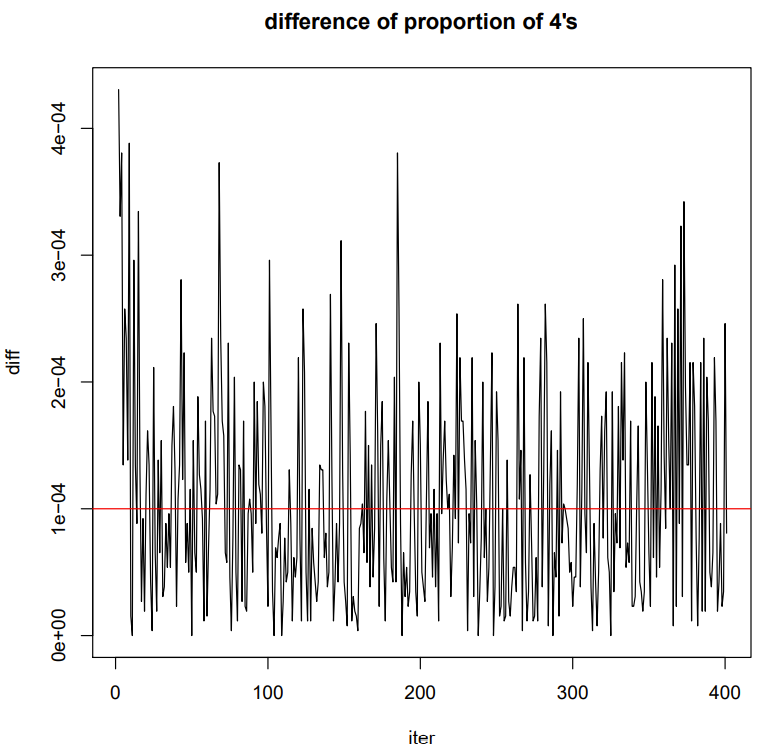} }%
    \qquad
    \subfloat[
    Difference between the proportion of frames with 8 white sites in the first frame across each iteration.]{\label{fig:diff8}\includegraphics[scale = 0.42]{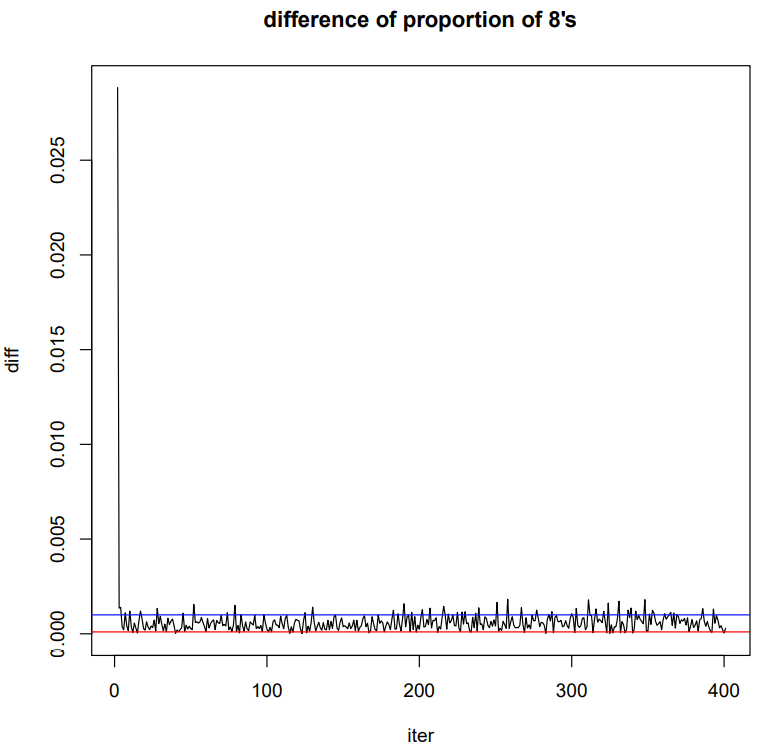} }
    \caption{Difference between the frequency of certain configurations within a matrix from one iteration to another, up to 400 iterations. The blue line represents a difference of $10^{-3}$ and the red one is $10^{-4}$.}%
    \label{fig:diff}%
    \end{figure}
    
    A total of 100 matrices with $510\times510$ sites were created and stored after 400 iterations of the resampling algorithm. This task was performed in approximately 9 hours using a machine from UFMG's Mathematics Department which has an Intel Xeon processor running at 3.8GHz and using 64GB of RAM. 
    
    The progression of these matrices throughout the iterations is shown in Figure \ref{fig:k}. It seems that the limiting distribution of this process tends to have the whole matrix become fire (except for water pixels). The PCN model is simply a snapshot of the process in the short term. Luckily, in the real world, other factors come into play to interrupt this process.
    
    Our interest lies in recovering the PCN tree representing this phenomenon, not recovering the image itself. We observed the difference in the frequencies of certain configurations from one iteration to the other to help decide when the matrices ``stabilized". Figure \ref{fig:diff} presents these results. The blue line represents a difference of $10^{-3}$ while the red line is $10^{-4}$. We consider the matrices to have met the stabilization criterion when the difference between iterations falls underneath the blue line. Therefore, the matrices appear to settle within just a few iterations.
    
    \begin{table}[H]
    \caption{Comparison between the probability of the site being fire given the context in the PCN $\hat{\mathcal{T}}$ and the estimated interval obtained from the bootstrap method. The lower bound (LB) corresponds to the 2.5\textsuperscript{th} percentile and the upper bound (UB) is the 97.5\textsuperscript{th} percentile.}
    \label{tab:pant}
    \centering
    \begin{tabular}{ c c c c c }
    \toprule
    \multirow{2}{*}{Context } &
    \multirow{2}{*}{PCN $\hat{\mathcal{T}}$} &
    \multicolumn{3}{c}{Interval Estimate } \\
    \cline{3-5}
    & & LB & Median & UB \\
    \midrule
    \begin{minipage}{.05\textwidth}
      \includegraphics[width=2em, height=2em]{Figures/1preto.PNG}
    \end{minipage}
    & 0.0431 & 0.0278 & 0.0386 & 0.0467 \\[1.5mm]
    \begin{minipage}{.05\textwidth}
      \includegraphics[width=2em, height=2em]{Figures/2pretos.PNG}
    \end{minipage}
    & 0.1230 & 0.0907 & 0.1111 & 0.1307  \\[1.5mm]
    \begin{minipage}{.05\textwidth}
      \includegraphics[width=2em, height=2em]{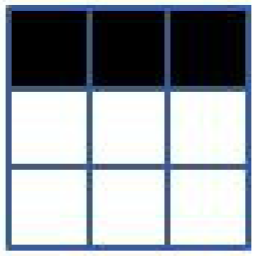}
    \end{minipage}
    & 0.2620 & 0.2375 & 0.2586 & 0.2814  \\[1.5mm]
    \begin{minipage}{.05\textwidth}
      \includegraphics[width=2em, height=2em]{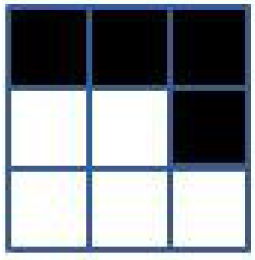}
    \end{minipage}
    & 0.5260 & 0.4793 & 0.5097 & 0.5322 \\[1.5mm]
    \begin{minipage}{.05\textwidth}
      \includegraphics[width=2em, height=2em]{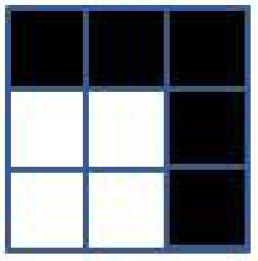}
    \end{minipage}
    & 0.8046 & 0.7396 & 0.7627 & 0.7851 \\[1.5mm]
    \begin{minipage}{.05\textwidth}
      \includegraphics[width=2em, height=2em]{Figures/6pretos.PNG}
    \end{minipage}
    & 0.8904 & 0.8821 & 0.8965 & 0.9109 \\[1.5mm]
    \begin{minipage}{.05\textwidth}
      \includegraphics[width=2em, height=2em]{Figures/7pretos.PNG}
    \end{minipage}
    & 0.9634 & 0.9616 & 0.9665 & 0.9712  \\[1.5mm]
    \begin{minipage}{.05\textwidth}
      \includegraphics[width=2em, height=2em]{Figures/8pretos.PNG}
    \end{minipage}
    & 0.9960 & 0.9952 & 0.9957 & 0.9962 \\[1.5mm]
    \begin{minipage}{.05\textwidth}
      \includegraphics[width=2em, height=2em]{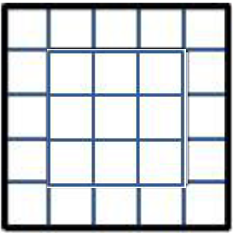}
    \end{minipage}
    & 0.0002 & 0.0001 & 0.0002 & 0.0002 \\[1.5mm]
    \begin{minipage}{.05\textwidth}
      \includegraphics[width=2em, height=2em]{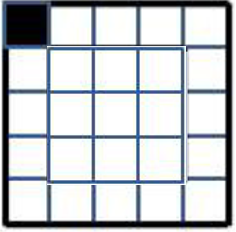}
    \end{minipage}
    & 0.0075 & 0.0024 & 0.0056 & 0.0103 \\[1.5mm]
    \begin{minipage}{.05\textwidth}
      \includegraphics[width=2em, height=2em]{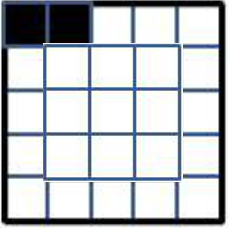}
    \end{minipage}
    & 0.0078 & 0.0014 & 0.0075 & 0.0154  \\[1.5mm]
    \begin{minipage}{.05\textwidth}
      \includegraphics[width=2em, height=2em]{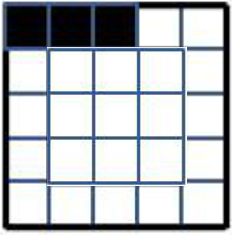}
    \end{minipage}
    & 0.0112 & 0.0019 & 0.0098 & 0.0224  \\[1.5mm]
    \begin{minipage}{.05\textwidth}
      \includegraphics[width=2em, height=2em]{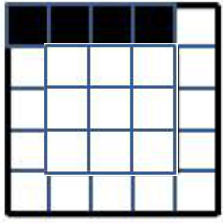}
    \end{minipage}
    & 0.0094 & 0.0032 & 0.0098 & 0.0225 \\[1.5mm]
     \begin{minipage}{.05\textwidth}
      \includegraphics[width=2em, height=2em]{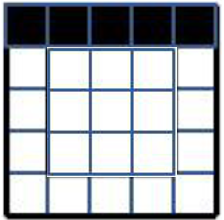}
    \end{minipage}
    & 0.0162 & 0.0000 & 0.0111 & 0.0356 \\[1.5mm]
        \begin{minipage}{.05\textwidth}
      \includegraphics[width=2em, height=2em]{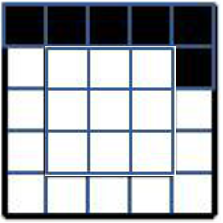}
    \end{minipage}
    & 0.0000 & 0.0000 & 0.0108 & 0.0395 \\[1.5mm]
    \begin{minipage}{.05\textwidth}
      \includegraphics[width=2em, height=2em]{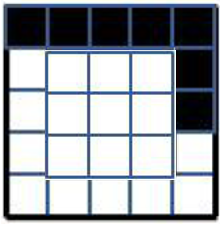}
    \end{minipage}
    & 0.0000 & 0.0000 & 0.0000 & 0.0645  \\[1.5mm]
    \begin{minipage}{.05\textwidth}
      \includegraphics[width=2em, height=2em]{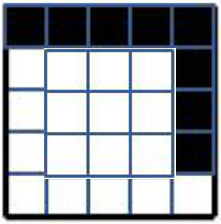}
    \end{minipage}
    & 0.0000 & 0.0000 & 0.0000 & 0.0909  \\[1.5mm]
    \begin{minipage}{.05\textwidth}
      \includegraphics[width=2em, height=2em]{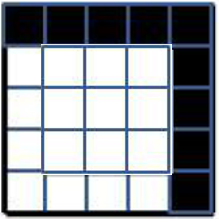}
    \end{minipage}
    & 0.0000 & 0.0000 & 0.0000 & 0.2129 \\[1.5mm]
    \begin{minipage}{.05\textwidth}
      \includegraphics[width=2em, height=2em]{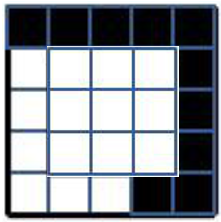}
    \end{minipage}
    & 0.0000 & 0.0000 & 0.0000 & 0.1833  \\[1.5mm]
    \begin{minipage}{.05\textwidth}
      \includegraphics[width=2em, height=2em]{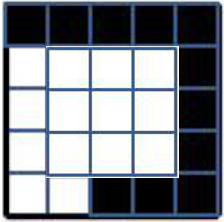}
    \end{minipage}
    & 0.0000 & 0.0000 & 0.0000 & 0.0000  \\[1.5mm]
    \begin{minipage}{.05\textwidth}
      \includegraphics[width=2em, height=2em]{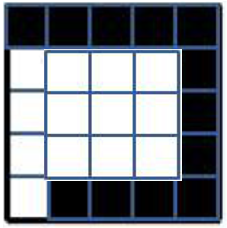}
    \end{minipage}
    & 0.0000 & 0.0000 & 0.0000 & 0.0000  \\[1.5mm]
     \begin{minipage}{.05\textwidth}
      \includegraphics[width=2em, height=2em]{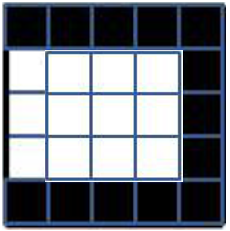}
    \end{minipage}
    & 0.0000 & 0.0000 & 0.0000 & 0.0000  \\[1.5mm]
    \begin{minipage}{.05\textwidth}
      \includegraphics[width=2em, height=2em]{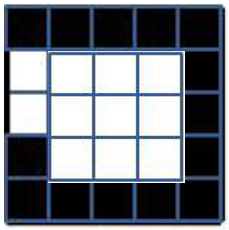}
    \end{minipage}
    & 0.0000 & 0.0000 & 0.0000 & 0.0000  \\[1.5mm]
    \begin{minipage}{.05\textwidth}
      \includegraphics[width=2em, height=2em]{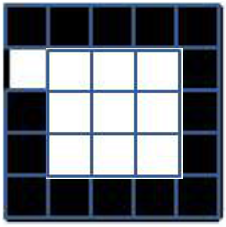}
    \end{minipage}
    & - & - & 0.0000 & -  \\[1.5mm]
    \bottomrule
    \end{tabular}
    \end{table}
    
    The estimated intervals were built based on the 2.5\textsuperscript{th} percentile, median, and 97.5\textsuperscript{th} percentile of the resample's conditional probabilities. Once again, we followed \cite{HYNDMAN1996}'s recommendation to use the median unbiased sample quantile estimator. 
    
    Table \ref{tab:pant} shows the results. In summary, the estimated intervals contain the conditional probability seen in $\hat{\mathcal{T}}$ for all contexts, except for the one with 5 sites of fire in the first-order neighborhood. In that case, the upper bound falls short by 0.0194. All intervals have a relatively small range of values, increasing the range as the frequency of the configurations decreases within the resample (and within the matrices belonging to the resample). The intervals whose lower bound, median and upper bound all equaled zero appeared, at most, 3 times within the matrices that contained those neighborhoods. Additionally, the neighborhood containing 15 fires in the second order, appeared in one matrix a single time. This is the reason why there is no upper bound or lower bound associated with it. This specific configuration was not observed in the Pantanal original matrix.
    

\section{Conclusion}
\label{sec:conclusion}

    The probabilistic context neighborhood (PCN) model proposed in this work offers a modeling alternative to studying the dependency structure of a discrete Markov process in a two-dimensional lattice, similar to the probabilistic context tree (PCT) model in the one-dimensional case \cite{CSISZAR2006a} when the size of the neighborhood may vary from one site to another. The tree structure of the PCN allows for easy interpretation of site dependencies, aiding understanding of data interactions (see Section \ref{sec:app}).
    
    The generalization to the multi-dimensional case was possible by replacing the likelihood with the pseudo-likelihood and the Bayesian information criterion (BIC) with the pseudo-Bayesian information criterion (PIC). In \cite{CSISZAR2006b}, the consistency of the PIC estimator for the candidate neighborhood of a site was proven, but an algorithm for the selection of the given estimator was not provided. The authors considered this task to be elusive. Since the PCN model sets a fixed frame neighborhood geometry, the cardinality of possible contexts can be calculated. The main advantage of the PCN model is the proposal of an algorithm that selects the optimal PCN tree without the burden of calculating the PIC score for all possibilities. 
    
    Our simulation study in Section \ref{sec:sim} showed our methodology's and algorithm's accuracy. The algorithm correctly recovered the PCN $\mathcal{T}_0$ that generated the sample in all scenarios.  In Section \ref{sec:app}, we showed the adequacy of our methodology for analyzing spatial data

    
   It is worth exploring several areas in further studies. One such area is the generalization of the model results to lattices in $\mathbb{Z}^d$, for $d>2$. Although the ease of interpretation given by visualizing the dependency tree in two dimensions is lost, we can think of other ways of presenting the dependency graph in higher dimensions. Additionally, the extension of this methodology to a more general graph structure (outside of a lattice) would be interesting to explore. In the case of maps, for example,  where the edges are also not random, the neighborhood can be defined by an adjacency matrix, and the extension is almost immediate. However, we must be careful when defining higher-order neighborhoods and boundaries, and the PCN algorithm provided here must be modified to deal with these definitions. In Another possible further study, we can include covariates in a regression model to better understand the dependence structure of each site. This way, variables like wind velocity and direction could improve the analysis of the Pantanal Fire data, for example. Besides, we can include a temporal component to understand the dynamic behavior of neighborhood dependence. Various options are available for carrying forward the research work discussed in this article.
Our final remark is that When dealing with continuous variables, our methodology can still be used, but with one condition: the values must first be divided into categories. This is because our main tool is counting finite configurations. In situations where categorization of the values is not feasible, an alternative methodology must be devised.
    \vspace{2em}

{\bf Acknowledgements} The authors thank CAPES and FAPEMIG for their financial support.

\bibliography{Thesis}

\end{document}